\documentclass[12pt]{article}
 
\usepackage{axodraw}

\def\comp{{\rm C}\llap{\vrule height7.1pt width1pt depth-.4pt\phantom t}}
\def\Fint{\rlap{$\Biggl\rfloor$}\Biggl\lceil}
\def\square{\kern1pt\vbox{\hrule height 1.2pt\hbox{\vrule width 1.2pt\hskip 3pt
   \vbox{\vskip 6pt}\hskip 3pt\vrule width 0.6pt}\hrule height 0.6pt}\kern1pt}

\begin{document}

\begin{titlepage}
 
\begin{flushright}
gr-qc/0607094 \\ SPIN-06/26, ITP-UU-06/30 \\ CRETE-06-15 \\ UFIFT-QG-06-07
\end{flushright}

\begin{center}
{\bf Two Loop Scalar Bilinears for Inflationary SQED}
\end{center}

\begin{center}
T. Prokopec$^*$
\end{center}

\begin{center}
\it{Institute for Theoretical Physics \& Spinoza Institute, Utrecht University\\
Leuvenlaan 4, Postbus 80.195, 3508 TD Utrecht, THE NETHERLANDS}
\end{center}

\begin{center}
N. C. Tsamis$^{\dagger}$
\end{center}

\begin{center}
\it{Department of Physics, University of Crete \\
GR-710 03 Heraklion, HELLAS}
\end{center}

\begin{center}
R. P. Woodard$^{\ddagger}$
\end{center}

\begin{center}
\it{Department of Physics, University of Florida \\
Gainesville, FL 32611, UNITED STATES}
\end{center}

\vspace{1cm}

\begin{center}
ABSTRACT
\end{center}
We evaluate the one and two loop contributions to the expectation 
values of two coincident and gauge invariant scalar bilinears in
the theory of massless, minimally coupled scalar quantum electrodynamics 
on a locally de Sitter background. One of these bilinears is the product 
of two covariantly differentiated scalars, the other is the product of 
two undifferentiated scalars. The computations are done using dimensional
regularization and the Schwinger-Keldysh formalism. Our results are
in perfect agreement with the stochastic predictions at this order.

\begin{flushleft}
\end{flushleft}

\begin{flushleft}
$^*$ e-mail: T.Prokopec@phys.uu.nl \\
$^{\dagger}$ e-mail: tsamis@physics.uoc.gr \\
$^{\ddagger}$ e-mail: woodard@phys.ufl.edu
\end{flushleft}

\end{titlepage}

\section{Introduction}

Quantum field theories which involve either massless, minimally coupled 
(MMC) scalars or gravitons have the propensity for vastly enhanced quantum
effects during inflation. These particles' combination of masslessness 
without classical conformal invariance results in prodigious particle 
production during inflation \cite{RPW1}. As more and more long wavelength, 
virtual quanta are ripped out of the vacuum, the MMC scalar and graviton 
field strengths grow like the logarithm of the inflationary scale factor.
This is evident even in free MMC scalar field theory \cite{VF,L,S},
\begin{equation}
\Bigl\langle \Omega \Bigl\vert \varphi^2(x) \Bigr\vert \Omega \Bigr\rangle_0 
= \frac{H^2}{4 \pi^2} \ln(a) + {\rm Divergent\ Constant} \; . \label{lowest}
\end{equation}
Of course interactions which involve undifferentiated MMC scalars or 
gravitons are correspondingly strengthened.

Powers of infrared logarithms like that in (\ref{lowest}) arise in the one 
particle irreducible (1PI) functions of a MMC scalar with a quartic
self-interaction \cite{OW1,OW2,BOW}. They occur as well in MMC scalar 
quantum electrodynamics (SQED) \cite{PTW1,PTW2,PW1,PW2} and in massless 
Yukawa theory \cite{PW3,GP}. The 1PI functions of pure gravity on de Sitter 
background show infrared logarithms \cite{TW1,TW2,TW3}, as do the 1PI
functions of Dirac + Einstein \cite{MW1,MW2}, and presumably gravity with any
other theory. Weinberg has recently drawn attention to their appearance 
in fixed-momentum correlation functions \cite{SW1,SW2}.

Infrared logarithms are fascinating because they can grow enough during a
long period of inflation to compensate for even the smallest coupling 
constant. However, what this really means is that perturbation theory 
breaks down, not necessarily that quantum effects become large. Deciding 
what actually happens requires a nonperturbative resummation technique. 

Starobinski\u{\i} has long argued that the nonperturbative evolution can be
followed using his stochastic reformulation of inflationary quantum field
theory \cite{AAS}. Probabilistic representations of inflationary cosmology 
have been much studied in order to understand initial conditions \cite{AV,NS}
and global structure \cite{GLM,LM} but we wish here to focus on 
Starobinski\u{\i}'s stochastic formulation as a wonderfully simple way of 
recovering secular effects in quantum field theory \cite{SJR,SNN,WV}. On de 
Sitter background the technique has recently been proven to exactly reproduce 
the leading infrared logarithms, at each order in perturbation theory, for 
any model of the form \cite{RPW2,TW4},
\begin{equation}
\mathcal{L} = -\frac12 \partial_{\mu} \varphi \partial_{\nu} \varphi
g^{\mu\nu} \sqrt{-g} - V(\varphi) \sqrt{-g} \; . \label{Staroform}
\end{equation}
Provided the potential in (\ref{Staroform}) is bounded below, Starobinski\u{\i} 
and Yokoyama have used the stochastic technique to give an explicit solution 
for the late time limit \cite{SY}.

Two important generalizations of Starobinski\u{\i}'s technique are necessary:
\begin{itemize}
\item{Apply it on de Sitter background to more complicated models which also 
show infrared logarithms, such as Yukawa \cite{PW3,GP}, SQED 
\cite{PTW1,PTW2,PW1,PW2} and quantum gravity 
\cite{TW1,TW2,TW3,MW1,MW2,SW1,SW2}; and}
\item{Apply it to a general inflationary and post-inflationary cosmological
background.}
\end{itemize}
This first step has already been taken for Yukawa \cite{MW3}. One of the
surprising features of the result is that {\it the ultraviolet cannot be
ignored, even at leading logarithm order.} Indeed, leading logarithm
results for coincident Green's functions can be ultraviolet divergent!
This stands in sharp contrast to the situation for models of the form 
(\ref{Staroform}). Aspects of quantum field theory are sufficiently 
counter-intuitive --- even in flat space! --- that it would be folly to
ignore the possibility of further surprises as the formalism is generalized 
to models with gauge symmetry and derivative interactions. It is therefore 
imperative to test putative generalizations of Starobinski\u{\i}'s technique
against explicit perturbative computations at the highest possible loop 
order. The purpose of this paper is to provide this sort of ``raw data''
for comparison with a leading logarithm formulation of SQED \cite{PTsW1}.

We report one and two loop results for the (Bunch-Davies) vacuum expectation 
values (VEV's) of two coincident, gauge invariant bilinears of the charged 
scalar field,
\begin{equation}
\Bigl\langle \Omega \Big\vert \Bigl(D_{\mu} \varphi(x)\Bigr)^* D_{\nu}
\varphi(x) \Bigr\vert \Omega \Bigr\rangle \qquad {\rm and} \qquad
\Bigl\langle \Omega \Bigl\vert \varphi^*(x) \varphi(x) \Bigr\vert \Omega
\Bigr\rangle \; . \label{twoVs}
\end{equation}
The Feynman rules are given section 2. Section 3 presents a key result
for the one loop self-mass-squared, which also serves to fix the divergent
parts of the conformal and scalar field strength counterterms at order
$e^2$. The first of the two VEV's in (\ref{twoVs}) is evaluated in section
4. Section 5 does the second. Our discussion comprises section 6, and the
less savory details are consigned to an Appendix.

\section{Feynman Rules}

The purpose of this section is to work out the Feynman rules for SQED 
in de Sitter conformal coordinates. We begin by reviewing the background 
geometry. Then the Lagrangian is given along with a precise definition 
of the renormalization parameters. Of course this allows one to read off 
the interactions in a straightforward manner. We next present the 
propagators. The section closes with a review of the Schwinger-Keldysh 
formalism which adapts the usual Feynman rules to give true expectation 
values rather than in-out amplitudes.

\subsection{Geometry in de Sitter}

We work in the conformal coordinate system of $D$-dimensional de 
Sitter space,
\begin{equation}
ds^2 = a^2 \Bigl( -d\eta^2 + d\vec{x} \cdot d\vec{x} \Bigr) \qquad
{\rm where} \qquad a(\eta) = -\frac1{H \eta} \; .
\end{equation}
Hence the metric is $g_{\mu\nu} = a^2 \eta_{\mu\nu}$, where $\eta_{\mu\nu}$
is the Minkowski metric. The affine connection for this background is,
\begin{equation}
\Gamma^{\rho}_{~\mu\nu} = H a \Bigl(\delta^{\rho}_{\mu} \delta^0_{\nu}
+ \delta^{\rho}_{\nu} \delta^0_{\mu} + \delta^{\rho}_{0} \eta_{\mu\nu} \Bigr) 
\; .
\end{equation}
For any geometry the scalar d`Alembertian is,
\begin{equation}
\square \equiv \frac1{\sqrt{-g}} \partial_{\mu} \Bigl( \sqrt{-g} 
g^{\mu\nu} \partial_{\nu} \Bigr) \; .
\end{equation}
And the vector d`Alembertian is defined by the relation,
\begin{equation}
\square_{\mu}^{~\nu} f_{\nu} \equiv g^{\rho\sigma} f_{\mu ; \rho \sigma}
\; ,
\end{equation}
where a semi-colin stands for covariant differentiation.

Propagators are most effectively expressed in terms of the length 
function, 
\begin{equation}
y(x;x') \equiv a a' \Bigl\{ \Vert \vec{x} \!-\! \vec{x}' \Vert^2 
- (\vert \eta \!-\! \eta'\vert \!-\! i \delta)^2 \Bigr\} \; .
\end{equation}
(Here $a \equiv a(\eta)$, $a' \equiv a(\eta')$, and the same convention is
employed throughout the paper. That is, $g_{\mu\nu} \equiv g_{\mu\nu}(x)$
and $g_{\mu\nu}' \equiv g_{\mu\nu}(x')$.)
In the limit that $\delta$ vanishes, $y(x;x')$ is related  to the invariant 
length $\ell(x;x')$ between the points $x^{\mu}$ and $x^{\prime \mu}$,
\begin{equation}
y(x;x') = 4 \sin^2\Bigl( \frac12 H \ell(x;x')\Bigr) \; .
\end{equation}
We often employ the following formula for acting $\square$ upon a
function of $y(x;x')$ which is analytic everywhere except possibly
at $y=0$,
\begin{eqnarray}
\lefteqn{\square f(y) = H^2 \Bigl\{ (4 y \!-\! y^2) f''(y) + D (2 \!-\! y) 
f'(y) \Bigr\} } \nonumber \\
& & \hspace{3cm} + {\rm Res}\Bigl[y^{\frac{D}2 -2} f\Bigr] \times \frac{4 
\pi^{\frac{D}2} H^{2-D}}{\Gamma(\frac{D}2 \!-\! 1)} \frac{i}{\sqrt{-g}} 
\delta^D(x \!-\! x') \; . \qquad \label{dAlem}
\end{eqnarray}
Here ${\rm Res}[F]$ stands for the residue of $F(y)$; that is, the coefficient
of $1/y$ in the Laurent expansion. 

Any scalar which depends upon $x^{\mu}$ and $x^{\prime \mu}$ can be considered
to be a function of $y(x;x')$. De Sitter invariant vector and tensor functions 
of $x^{\mu}$ and $x^{\prime \mu}$ can be represented by including the metric 
and just the first two derivatives of $y(x;x')$ \cite{AJ,KW},
\begin{eqnarray}
\frac{\partial y}{\partial x^{\mu}} = H a \Bigl(y \delta^0_{\mu} + 2 a' H
\Delta x_{\mu}\Bigr) \qquad , \qquad \frac{\partial y}{\partial x^{\prime \nu}}
= H a' \Bigl(y \delta^0_{\nu} - 2 a H \Delta x_{\nu}\Bigr) \; , \label{base1} \\
\frac{\partial^2 y}{\partial x^{\mu} \partial x^{\prime \nu}} = H^2 a a'
\Bigl(y \delta_{\mu}^0 \delta_{\nu}^0 - 2 a \delta_{\mu}^0 H \Delta x_{\nu}
+ 2 a' H \Delta x_{\mu} \delta_{\nu}^0 - 2 \eta_{\mu\nu}\Bigr) \; .\label{base2}
\end{eqnarray}
Here $\Delta x_{\mu} \equiv \eta_{\mu\nu} (x \!-\! x')^{\nu}$. Contracting 
any two of these basis tensors, on either primed or unprimed indices, 
produces a linear combination of the basis tensors \cite{AJ,KW},
\begin{eqnarray}
g^{\mu\rho}(x) \frac{\partial y}{\partial x^{\mu}} \frac{\partial y}{\partial 
x^{\rho}} & = & H^2 (4 y - y^2) \; , \label{ID1} \\
g^{\mu\rho}(x) \frac{\partial y}{\partial x^{\mu}} \frac{\partial^2 y}{\partial 
x^{\rho} \partial x^{\prime \nu}} & = & H^2 (2 - y) \frac{\partial y}{\partial
x^{\prime \nu}} \; , \label{ID2} \\
g^{\mu\rho}(x) \frac{\partial^2 y}{\partial x^{\mu} \partial x^{\prime \nu}} 
\frac{\partial^2 y}{\partial x^{\rho} \partial x^{\prime \sigma}} & = & 
4 H^4 g'_{\nu\sigma} - H^2 \frac{\partial y}{\partial x^{\prime \nu}} 
\frac{\partial y}{\partial x^{\prime \sigma}} \; . \label{ID3}
\end{eqnarray}
The same is true for covariant differentiation \cite{AJ,KW},
\begin{equation}
\Bigl(\frac{\partial y}{\partial x^{\mu}}\Bigr)_{;\nu} = H^2 (2 \!-\! y)
g_{\mu\nu}(x) \; .
\end{equation}

\subsection{Renormalizing SQED}

The bare Lagrangian of SQED is,
\begin{eqnarray}
\lefteqn{\mathcal{L} = - \Bigl(\partial_{\mu} \!-\! i e_0 A_{\mu}\Bigr)
\varphi^* \Bigl(\partial_{\nu} \!+\! i e_0 A_{\nu} \Bigr) \varphi g^{\mu\nu} 
\sqrt{-g} - \xi_0 \varphi^* \varphi R \sqrt{-g} } \nonumber \\
& & \hspace{4.5cm} - \frac14 \lambda_0 (\varphi^* \varphi)^2 \sqrt{-g} 
-\frac14 F_{\rho\sigma} F_{\mu\nu} g^{\rho\mu} g^{\sigma\nu} \sqrt{-g} 
\; . \qquad
\end{eqnarray}
Here $e_0$, $\xi_0$ and $\lambda_0$ are the bare couplings. No bare mass
is required because we study massless SQED, and mass is multiplicatively 
renormalized in dimensional regularization. Of course we also study minimally 
coupled SQED, but there is no similar relation for the conformal coupling in 
dimensional regularization. Hence $\xi_0$ must appear as a counterterm, even
though its renormalized value is zero.

The bare fields are expressed as usual in terms of the renormalized 
fields,
\begin{equation}
\varphi \equiv \sqrt{Z_2} \varphi_R \qquad {\rm and} \qquad A_{\mu} \equiv
\sqrt{Z_3} A_{R\mu} \; .
\end{equation}
In terms of the renormalized fields the Lagrangian takes the form,
\begin{eqnarray}
\lefteqn{\mathcal{L} \!=\! -Z_2 \Bigl(\partial_{\mu} \!\!-\! i \sqrt{Z_3} e_0 
A_{R \mu}\Bigr) \varphi_R^* \Bigl(\partial_{\nu} \!\!+\! \sqrt{Z_3} i e_0 
A_{R \nu} \Bigr) \varphi_R g^{\mu\nu} \sqrt{\!-g} \!-\! Z_2 \xi_0 \varphi_R^* 
\varphi_R R \sqrt{\!-g} } \nonumber \\
& & \hspace{3cm} - \frac14 Z_2^2 \lambda_0 (\varphi_R^* \varphi_R)^2 
\sqrt{-g} -\frac14 Z_3 F_{R \rho\sigma} F_{R \mu\nu} g^{\rho\mu} g^{\sigma\nu} 
\sqrt{-g} \; . \qquad
\end{eqnarray}
The various bare coupling constants can be expressed as follows in terms
of renormalized couplings and renormalization parameters,
\begin{equation}
\sqrt{Z_3} e_0 = e + 0 \qquad , \qquad Z_2 \xi_0 = 0 + \delta \xi \qquad 
{\rm and} \qquad Z_2^2 \lambda_0 = 0 + \delta \lambda \; .
\end{equation}
Note that we have chosen to make the renormalized 4-scalar coupling zero, 
as we are free to do. Defining the field strength renormalizations as usual,
\begin{equation}
Z_2 \equiv 1 + \delta Z_2 \qquad {\rm and} \qquad Z_3 \equiv 1 +
\delta Z_3 \; ,
\end{equation}
and dropping the now-redundant subscript $R$ on the renormalized fields, we
at length reach the form,
\begin{eqnarray}
\lefteqn{\mathcal{L} = - \Bigl(\partial_{\mu} \!-\! i e A_{\mu}\Bigr)
\varphi^* \Bigl(\partial_{\nu} \!+\! i e A_{\nu} \Bigr) \varphi g^{\mu\nu} 
\sqrt{-g} -\frac14 F_{\rho\sigma} F_{\mu\nu} g^{\rho\mu} g^{\sigma\nu} 
\sqrt{-g} } \nonumber \\
& & \hspace{1cm} - \delta Z_2 \Bigl(\partial_{\mu} \!-\! i e A_{\mu}\Bigr) 
\varphi^* \Bigl(\partial_{\nu} \!+\! i e A_{\nu} \Bigr) \varphi g^{\mu\nu} 
\sqrt{-g} - \delta \xi \varphi^* \varphi R \sqrt{-g} \nonumber \\
& & \hspace{4cm} - \frac14 \delta \lambda (\varphi^* \varphi)^2 \sqrt{-g} 
-\frac14 \delta Z_3 F_{\rho\sigma} F_{\mu\nu} g^{\rho\mu} g^{\sigma\nu} 
\sqrt{-g} \; . \qquad
\end{eqnarray}
For the computation reported here we shall not require the interactions 
proportional to $\delta \lambda$ and $\delta Z_3$ because they do not 
contribute to either VEV in (\ref{twoVs}) at order $e^2$.

\subsection{Propagators}

The scalar propagator obeys,
\begin{equation}
\sqrt{-g} \square i\Delta(x;x') = i \delta^D(x \!-\! x') \; . \label{scprop}
\end{equation}
It has long been known that there is no de Sitter invariant solution
\cite{AF}. If one elects to break de Sitter invariance while preserving 
homogeneity and isotropy --- this is known as the ``E(3)'' vacuum \cite{BA} 
--- the minimal solution takes the form \cite{OW1,OW2},
\begin{equation}
i\Delta(x;x') = A(y) + k \ln(a a') \qquad {\rm where} \qquad
k \equiv \frac{H^{D-2}}{(4\pi)^{\frac{D}2}} \frac{\Gamma(D\!-\!1)}{
\Gamma(\frac{D}2)} \; .
\end{equation}
The de Sitter invariant function $A(y)$ is \cite{OW2},
\begin{eqnarray}
\lefteqn{A(y) \equiv \frac{H^{D-2}}{(4\pi)^{\frac{D}2}} \Biggl\{
\frac{\Gamma(\frac{D}2 \!-\! 1)}{\frac{D}2 \!-\! 1} \Bigl(\frac{4}{y}\Bigr)^{
\frac{D}2 -1} \!+\! \frac{\Gamma(\frac{D}2 \!+\! 1)}{\frac{D}2 \!-\! 2} 
\Bigl(\frac{4}{y} \Bigr)^{\frac{D}2-2} \!-\! \pi \cot\Bigl(\frac{\pi D}2\Bigr) 
\frac{\Gamma(D \!-\! 1)}{\Gamma(\frac{D}2)} } \nonumber \\
& & \hspace{1cm} + \sum_{n=1}^{\infty} \Biggl[\frac1{n} \frac{\Gamma(n \!+\! 
D \!-\! 1)}{\Gamma(n \!+\! \frac{D}2)} \Bigl(\frac{y}4 \Bigr)^n \!\!\!\! - 
\frac1{n \!-\! \frac{D}2 \!+\! 2} \frac{\Gamma(n \!+\!  \frac{D}2 \!+\! 1)}{
\Gamma(n \!+\! 2)} \Bigl(\frac{y}4 \Bigr)^{n - \frac{D}2 +2} \Biggr] \Biggr\} . 
\qquad \label{A}
\end{eqnarray}

Expression (\ref{A}) may seem daunting but it is actually simple to use because
the infinite sum vanishes in $D\!=\!4$, and the terms of this sum go like
higher and higher powers of $y(x;x')$. Hence the infinite sum can only 
contribute when multiplied by a divergent term, and even then only the first
few terms can contribute. It turns out that most computations in this paper
require only the derivative, $A'(y)$, expanded to the following order,
\begin{equation}
A'(y) = -\frac14 \Gamma\Bigl(\frac{D}2\Bigr) \frac{H^{D-2}}{(4\pi)^{\frac{D}2}}
\Biggl\{\Bigl(\frac{4}{y}\Bigr)^{\frac{D}2} \!+\! \frac{D}2 \Bigl(\frac{4}{y} 
\Bigr)^{\frac{D}2-1} \!+\! O(D\!-\!4) \Biggr\} . \qquad \label{Aexp}
\end{equation}
We also need the exact coincidence limit,
\begin{equation}
A(0) = -k \pi \cot\Bigl(\frac{\pi D}2\Bigr) \; .
\end{equation}
Hence the coincidence limit of the scalar propagator is,
\begin{equation}
\lim_{x' \rightarrow x} i\Delta(x;x') = A(0) + 2 k \ln(a) \; .
\end{equation}

The VEV's we seek (\ref{twoVs}) are both of gauge invariant operators,
so it does not matter what gauge we use. The photon propagator has been
worked out in a variety of de Sitter invariant \cite{AJ} and noninvariant
\cite{KW} gauges. The calculations of this paper happen to simplify greatly 
in Lorentz gauge,
\begin{equation}
\partial_{\rho} \Bigl\{ \sqrt{-g} g^{\rho\mu} i\Bigl[\mbox{}_{\mu} 
\Delta_{\nu}\Bigr](x;x') \Bigr\} = 0 = \partial'_{\sigma} \Bigl\{ \sqrt{-g'} 
g^{\prime \sigma\nu} i\Bigl[\mbox{}_{\mu} \Delta_{\nu}\Bigr](x;x') \Bigr\}
\; . \label{gauge}
\end{equation}
The general form of the photon propagator in any de Sitter invariant gauge 
such as this is \cite{AJ,KW},
\begin{equation}
i\Bigl[\mbox{}_{\mu} \Delta_{\nu}\Bigr](x;x') = B(y) \frac{\partial^2 y}{
\partial x^{\mu} \partial x^{\prime \nu}} + C(y) \frac{\partial y}{\partial
x^{\mu}} \frac{\partial y}{\partial x^{\prime \nu}} \; . \label{genform}
\end{equation}
For Lorentz gauge (\ref{gauge}) the two functions $B(y)$ and $C(y)$ can be 
expressed as follows in terms of a single function $\gamma(y)$ \cite{AJ,TW5},
\begin{eqnarray}
B(y) & = & \frac1{4 (D\!-\!1) H^2} \Bigl[ -(4y \!-\! y^2) \gamma'(y) 
- (D\!-\!1) (2 \!-\! y) \gamma(y)\Bigr] \label{Brel} \; , \\
C(y) & = & \frac1{4 (D\!-\!1) H^2} \Bigl[ +(2 \!-\! y) \gamma'(y) 
- (D\!-\!1) \gamma(y)\Bigr] \label{Crel} \; .
\end{eqnarray}

The gauge condition (\ref{gauge}) is obeyed for any function $\gamma(y)$.
What fixes $\gamma(y)$ is the equation for the photon propagator \cite{TW5},
\begin{equation}
\sqrt{-g} \Bigl(\square_{\mu}^{~\rho} - R_{\mu}^{~\rho} \Bigr)
i\Bigl[\mbox{}_{\rho} \Delta_{\nu}\Bigr](x;x') = i g_{\mu\nu} \delta^D(x\!-\!x)
+ \sqrt{-g} \partial_{\mu} \partial'_{\nu} i\Delta(x;x') \; . \label{phprop}
\end{equation}
The unique de Sitter invariant solution for $\gamma(y)$ is \cite{TW5},
\begin{eqnarray}
\lefteqn{\gamma(y) \equiv \frac12 (D\!-\!1) \frac{H^{D-2}}{(4\pi)^{\frac{D}2}} 
\Biggl\{\frac{\Gamma(\frac{D}2)}{\frac{D}2 \!-\! 1} \Bigl(\frac{4}{y}\Bigr)^{
\frac{D}2 -1}  + \frac1{D\!-\!3} \sum_{n=0}^{\infty} \Biggl[\frac{\Gamma(n 
\!+\! D \!-\! 1)}{\Gamma(n \!+\! \frac{D}2 \!+\! 1)} \Bigl(\frac{y}4\Bigr)^n } 
\nonumber \\
& & \hspace{1.5cm} \times (n\!+\!1) \Bigl[\psi\Bigl(2 \!-\!\frac{D}2\Bigr) 
\!-\! \psi\Bigl(\frac{D}2 \!-\! 1\Bigr) + \psi\Bigl(n \!+\! D\!-\! 1\Bigr) 
\!-\! \psi\Bigl(n \!+\! 2\Bigr)\Bigr] \nonumber \\
& & - \frac{\Gamma(n \!+\! \frac{D}2 \!+\! 1)}{\Gamma(n \!+\! 3)} 
\Bigl(\frac{y}4\Bigr)^{n+3-\frac{D}2} \Bigl(n\!+\!3 \!-\! \frac{D}2\Bigr) 
\nonumber \\
& & \hspace{1.5cm} \times \Bigl[\psi\Bigl(2 \!-\!\frac{D}2 \Bigr) \!-\! 
\psi\Bigl(\frac{D}2 \!-\! 1\Bigr) + \psi\Bigl(n \!+\! \frac{D}2 \!+\! 1\Bigr) 
\!-\! \psi\Bigl(n \!+\! 4 \!-\! \frac{D}2\Bigr)\Bigr] \Biggr] \Biggr\} . 
\qquad \label{gamexp}
\end{eqnarray}
Here the symbol ``$\psi(z)$'' stands for the polygamma function,
\begin{equation}
\psi(z) \equiv \frac{d}{dz} \ln\Bigl(\Gamma(z)\Bigr) \; .
\end{equation}

As with the scalar propagator, we do not require the full complexity
of $\gamma(y)$. For most computations in this paper the following 
expansions suffice,
\begin{eqnarray}
\gamma(y) & = & \frac12 (D\!-\!1) \Gamma\Bigl(\frac{D}2\Bigr)
\frac{H^{D-2}}{(4\pi)^{\frac{D}2}} \Biggl\{ \frac2{D \!-\!2} \Bigl(\frac4{y}
\Bigr)^{\frac{D}2 -1} \nonumber \\
& & \hspace{5cm} + \frac{\ln(\frac{y}4)}{(1 \!-\! \frac{y}4)^2} + 
\frac1{1 \!-\! \frac{y}4} + O(D\!-\!4) \Biggr\} , \qquad \label{gamnor} \\
\gamma'(y) & = & -\frac18 (D\!-\!1) \Gamma\Bigl(\frac{D}2\Bigr)
\frac{H^{D-2}}{(4\pi)^{\frac{D}2}} \Biggl\{ \Bigl(\frac4{y}\Bigr)^{\frac{D}2}
+ \Bigl[\frac{D}2 \!-\! 2 \!-\! \frac2{D\!-\!2}\Bigr] \Bigl(\frac4{y}\Bigr)^{
\frac{D}2-1} \nonumber \\
& & \hspace{3cm} - \frac{2 \ln(\frac{y}4)}{(1 \!-\! \frac{y}4)^3} - \frac2{(1 
\!-\! \frac{y}4)^2} - \frac1{1 \!-\! \frac{y}4} + O(D\!-\!4) \Biggr\} ,
\qquad \label{gamp} \\
B(y) & = & \frac14 \Gamma\Bigl(\frac{D}2\Bigr) \frac{H^{D-4}}{(4\pi)^{
\frac{D}2}} \Biggl\{-\frac2{D \!-\!2} \Bigl(\frac4{y} \Bigr)^{\frac{D}2 -1} 
\nonumber \\
& & \hspace{3cm} - \frac{\ln(\frac{y}4)}{(1 \!-\! \frac{y}4)^2} -
\frac{2 \ln(\frac{y}4)}{1 \!-\! \frac{y}4} - \frac1{1 \!-\! \frac{y}4} + 
O(D\!-\!4) \Biggr\} , \qquad \label{Bexp} \\
C(y) & = & \frac1{16} \Gamma\Bigl(\frac{D}2\Bigr) \frac{H^{D-4}}{(4\pi)^{
\frac{D}2}} \Biggl\{- \Bigl(\frac4{y} \Bigr)^{\frac{D}2} - \Bigl[\frac{D}2 
\!+\! \frac{2}{D\!-\!2} \Bigr] \Bigl(\frac4{y} \Bigr)^{\frac{D}2 -1} 
\nonumber \\
& & \hspace{1cm} - \frac{2 \ln(\frac{y}4)}{(1 \!-\! \frac{y}4)^3} -
\frac{2 \ln(\frac{y}4)}{(1 \!-\! \frac{y}4)^2} - \frac2{(1 \!-\! \frac{y}4)^2} 
- \frac3{1 \!-\! \frac{y}4} + O(D\!-\!4) \Biggr\} . \qquad \label{Cexp}
\end{eqnarray}
We also need the exact coincidence limit,
\begin{eqnarray}
\lefteqn{\gamma(0) = \frac12 \Bigl(\frac{D\!-\!1}{D\!-\!3}\Bigr) 
\frac{H^{D-2}}{(4\pi)^{\frac{D}2}} \frac{\Gamma(D \!-\! 1)}{\Gamma(\frac{D}2
\!+\!1)} } \nonumber \\
& & \hspace{4cm} \times \Bigl[\psi\Bigl(2 \!-\!\frac{D}2\Bigr) \!-\! 
\psi\Bigl(\frac{D}2 \!-\! 1\Bigr) + \psi(D\!-\! 1) \!-\! \psi(2)\Bigr] \Biggr\}
. \qquad
\end{eqnarray}
The coincidence limit of the photon propagator follows from this, combined 
with expressions (\ref{genform}), (\ref{Brel}-\ref{Crel}) and 
(\ref{base1}-\ref{base2}),
\begin{equation}
\lim_{x' \rightarrow x} i\Bigl[\mbox{}_{\mu} \Delta_{\nu}\Bigr](x;x') = 
\gamma(0) g_{\mu\nu} \; .
\end{equation}

\subsection{The Schwinger-Keldysh formalism}

The Feynman rules we have just presented would suffice for computing the
matrix element of any operator between the state which is free vacuum in the
asymptotic past and the state which is free vacuum in the asymptotic future.
These in-out matrix elements are well adapted to scattering experiments in
flat space but they do not correspond to observations that can be performed
in de Sitter. There is no S-matrix in de Sitter \cite{EW,AS}. In fact the
vast expansion of spacetime in the infinite future means that in-out matrix 
elements even diverge off-shell \cite{TW6}. 

The physics reason behind the math problem is that free vacuum is an 
infinitely poor guess for the state after an endless history of inflationary
particle production. Indeed, the fact that we do not know what becomes of
the state in the infinite future is the whole reason the computation is
interesting! Under these circumstances, a more realistic theoretical 
question is, what happens if universe is released in free vacuum (or some 
other state) at a fixed time and then left to evolve as it will? One can 
follow this evolution by computing the expectation values of time-dependent,
local operators in the presence of the Heisenberg state. 

Of course expectation values of the sort just described could be computed
canonically. However, the canonical formalism is extremely cumbersome because 
it does not reflect the underlying spacetime symmetries in a manifest way. 
Schwinger long ago modified the ordinary Feynman rules to provide a manifestly 
covariant procedure for computing expectation values \cite{JS,M,BM,K}. Many 
excellent reviews of this formalism exist \cite{CSHY,J,CH,SW1} so we will 
merely give the key identity which relates it to the operator formalism 
\cite{FW},
\begin{eqnarray}
\lefteqn{\Bigl\langle \Psi \Bigl\vert \overline{T}^*\Bigl(\mathcal{O}_2[
\phi]\Bigr) T^*\Bigl(\mathcal{O}_1[\phi]\Bigr) \Bigr\vert \Psi 
\Bigr\rangle = \Fint [d\phi_+] [d\phi_-] \, \delta\Bigl[\phi_-(\ell) 
\!-\! \phi_+(\ell)\Bigr] } \nonumber \\
& & \hspace{1.5cm} \times \mathcal{O}_2[\phi_-] \mathcal{O}_1[\phi_+] 
\Psi^*[\phi_-(s)] e^{i \int_s^{\ell} dt \Bigl\{L[\phi_+(t)] - 
L[\phi_-(t)]\Bigr\}} \Psi[\phi_+(s)] \; . \qquad \label{fund}
\end{eqnarray}
Expression (\ref{fund}) is formulated in the context of a scalar field 
$\phi(x)$ whose Lagrangian (the space integral of the Lagrange density) at 
time $t$ is $L[\phi(t)]$. The left hand side of (\ref{fund}) is the canonical 
expectation value of the product of two arbitrary operators, 
$\mathcal{O}_1[\phi]$ and $\mathcal{O}_2[\phi]$, in the presence of a 
Heisenberg state $\vert \Psi\rangle$ whose wave functional in terms of the 
fields at the starting time $t=s$ is $\Psi[\phi(s)]$. We have time-ordered 
$\mathcal{O}_1[\phi]$ and anti-time-ordered $\mathcal{O}_2[\phi]$. The right 
hand side gives the functional integral representation of this canonical 
expectation value. The last time $t=\ell$ can be any point in the future of 
the latest operator in either $\mathcal{O}_1[\phi]$ or $\mathcal{O}_2[\phi]$.

It is important to remember that there is only one type of operator for
each of the fundamental fields. The $\comp$-number functions $\phi_{\pm}$ 
which appear on the right hand side of (\ref{fund}) are merely dummy variables 
in a convenient functional representation for the canonical expectation value 
on the left hand side. From this functional representation we can read off the 
following rules:
\begin{itemize}
\item{Each endpoint of each propagator carries a ``polarity'' which can be 
either ``$+$'' or ``$-$'';}
\item{External lines emanating from operators in $\mathcal{O}_1[\phi]$ are 
``$+$'', whereas those from operators in $\mathcal{O}_2[\phi]$ are ``$-$'';}
\item{Any given vertex is either all ``$+$'' or all ``$-$'';}
\item{The ``$+$'' vertices are identical to those of the in-out formalism,
 whereas the ``$-$'' vertices have the opposite sign;}
\item{The state wave functional $\Psi[\phi(s)]$ is assumed to be free vacuum,
plus possible perturbative corrections which would show up as interactions on 
the initial value surface; and}
\item{For SQED in de Sitter, the various polarities of the propagators are
obtained from the Feynman propagators of the previous subsection by making 
the following replacements for the length function $y(x;x')$ \cite{TW7},
\begin{eqnarray}
y_{\scriptscriptstyle ++}(x;x') & \longrightarrow a a' \Bigl\{ \Vert \vec{x} 
\!-\! \vec{x}' \Vert^2 - (\vert \eta \!-\! \eta'\vert \!-\! i \delta)^2 
\Bigr\} \; , \label{Y++} \\
y_{\scriptscriptstyle +-}(x;x') & \longrightarrow a a' \Bigl\{ \Vert \vec{x} 
\!-\! \vec{x}' \Vert^2 - (\eta \!-\! \eta' \!+\! i \delta)^2 \Bigr\} \; , 
\label{y+-} \\
y_{\scriptscriptstyle -+}(x;x') & \longrightarrow a a' \Bigl\{ \Vert \vec{x} 
\!-\! \vec{x}' \Vert^2 - (\eta \!-\! \eta' \!-\! i \delta)^2 \Bigr\} \; , 
\label{y-+} \\
y_{\scriptscriptstyle --}(x;x') & \longrightarrow a a' \Bigl\{ \Vert \vec{x} 
\!-\! \vec{x}' \Vert^2 - (\vert \eta \!-\! \eta'\vert \!+\! i \delta)^2 
\Bigr\} \; . \label{y--}
\end{eqnarray}}
\end{itemize}
It is worth calling attention to some consequences of these rules,
\begin{itemize}
\item{To each $N$-point 1PI function of the in-out formalism there 
correspond $2^N$ 1PI N-point functions in the Schwinger-Keldysh formalism;}
\item{The absence of mixed polarity counterterms means that no mixed
polarity 1PI functions harbor primitive divergences \cite{FW}; and}
\item{The propagators of mixed polarity obey the homogeneous analogues
of (\ref{scprop}) and (\ref{phprop}).}
\end{itemize}

\section{One Loop Self-Mass-Squared}

The purpose of this section is to derive a key result for the one loop
self-mass-squared. By renormalizing this quantity we also determine the order 
$e^2$ contributions to the renormalization parameters $\delta \xi$
and $\delta Z_2$.

The three diagrams which contribute to the $++$ polarization of the one 
loop self-mass-squared are depicted in Fig.~1.
\begin{center}
\begin{picture}(390,100)(0,0)
\DashArrowLine(40,50)(10,50){3}
\DashArrowLine(110,50)(40,50){3}
\DashArrowLine(140,50)(110,50){3}
\PhotonArc(75,50)(35,0,180){3}{12}
\Vertex(40,50){3}
\Text(40,37)[b]{$x$}
\Vertex(110,50){3}
\Text(110,37)[b]{$x'$}
\Text(160,45)[b]{\Large $+$}
\DashArrowLine(210,50)(180,50){3}
\DashArrowLine(240,50)(210,50){3}
\PhotonArc(210,67)(15,-90,270){2}{12}
\Vertex(210,50){3}
\Text(210,37)[b]{$x$}
\Text(260,45)[b]{\Large $+$}
\DashArrowLine(310,50)(280,50){3}
\DashArrowLine(340,50)(310,50){3}
\Vertex(310,50){3}
\Text(311,44)[b]{\LARGE $\times$}
\Text(310,37)[b]{$x$}
\end{picture}
\\ {\rm Fig.~1: One loop contributions to $-i M^2(x;x')$.}
\end{center}
Of course the left-hand diagram is the most difficult. In an arbitrary gauge 
it would consist of four terms. However, one of the nice things about Lorentz
gauge (\ref{gauge}) is that each of these terms makes the same contribution,
\begin{eqnarray}
\lefteqn{- e^2 \sqrt{-g} g^{\mu\rho} \sqrt{-g'} g^{\prime \nu \sigma}
i\Bigl[\mbox{}_{\rho} \Delta_{\sigma}\Bigr](x;x') \partial_{\mu} 
\partial'_{\nu} i\Delta(x;x') } \nonumber \\
& & - e^2 \partial_{\mu} \Bigl\{\sqrt{-g} g^{\mu\rho} \sqrt{-g'} g^{\prime 
\nu \sigma} i\Bigl[\mbox{}_{\rho} \Delta_{\sigma}\Bigr](x;x') \partial'_{\nu} 
i\Delta(x;x') \Bigr\} \nonumber \\
& & - e^2 \partial'_{\nu} \Bigl\{\sqrt{-g} g^{\mu\rho} \sqrt{-g'} g^{\prime 
\nu \sigma} i\Bigl[\mbox{}_{\rho} \Delta_{\sigma}\Bigr](x;x') \partial_{\mu} 
i\Delta(x;x') \Bigr\} \nonumber \\
& & - e^2 \partial_{\mu} \partial'_{\nu} \Bigl\{\sqrt{-g} g^{\mu\rho} 
\sqrt{-g'} g^{\prime \nu \sigma} i\Bigl[\mbox{}_{\rho} \Delta_{\sigma}
\Bigr](x;x') i\Delta(x;x') \Bigr\} \nonumber \\
& & \hspace{2.5cm} = -4 e^2 \sqrt{-g} g^{\mu\rho} \sqrt{-g'} g^{\prime \nu
\sigma} i\Bigl[\mbox{}_{\rho} \Delta_{\sigma}\Bigr](x;x') \partial_{\mu} 
\partial'_{\nu} i\Delta(x;x') \; . \qquad
\end{eqnarray}

The next step is to take the derivatives,
\begin{equation}
\partial_{\mu} \partial'_{\nu} i\Delta(x;x') = A'(y) \frac{\partial^2 y}{
\partial x^{\mu} \partial^{\prime \nu}} + A''(y) \frac{\partial y}{\partial
x^{\mu}} \frac{\partial y}{\partial x^{\prime \nu}} + \delta^0_{\mu}
\delta^0_{\nu} \frac{i}{a^{D-2}} \delta^D(x\!-\!x') \; ,
\end{equation}
and carry out the tensor contractions with the aid of identities 
(\ref{ID1}-\ref{ID3}). One then substitutes relations (\ref{Brel}) and 
(\ref{Crel}). The analysis is,
\begin{eqnarray}
\lefteqn{-4 e^2 \sqrt{-g} g^{\mu\rho} \sqrt{-g'} g^{\prime \nu \sigma}
i\Bigl[\mbox{}_{\rho} \Delta_{\sigma}\Bigr](x;x') \partial_{\mu} 
\partial'_{\nu} i\Delta(x;x') } \nonumber \\
& & = -4 e^2 \sqrt{-g} g^{\mu\rho} \sqrt{-g'} g^{\prime \nu \sigma} 
\Biggl\{ B \frac{\partial^2 y}{\partial x^{\rho} \partial x^{\prime \sigma}}
+ C \frac{\partial y}{\partial x^{\rho}} \frac{\partial y}{\partial x^{\prime
\sigma}} \Biggr\} \partial_{\mu} \partial'_{\nu} i\Delta(x;x') \; , \\
& & = - 4 e^2 H^4 \sqrt{-g} \sqrt{-g'} \Biggl\{ (4y \!-\!y^2) A'' \Bigl[ 
(2\!-\!y) B + (4y \!-\! y^2) C \Bigr] + 4 D B A' \nonumber \\
& & \hspace{2cm} - (4y \!-\! y^2) A' \Bigl[B - (2 \!-\! y) C\Bigr] \Biggr\} 
+ i 4 e^2 \gamma(0) \sqrt{-g} \delta^D(x \!-\!x') \; , \qquad \\
& & = 4 e^2 H^2 \sqrt{-g} \sqrt{-g'} \Biggl\{ (4 y \!-\! y^2) \Bigl[A'' 
\gamma \!+\! A' \gamma'\Bigr] + D (2 \!-\! y) A' \gamma \Biggr\} \nonumber \\
& & \hspace{7cm} + i 4 e^2 \gamma(0) \sqrt{-g} \delta^D(x \!-\! x') \; .
\qquad \label{lastex}
\end{eqnarray}

By comparing with (\ref{dAlem}) we can recognize (\ref{lastex}) as the scalar 
d`Alembertian acting upon the indefinite integral of $A'(y) \gamma(y)$. 
Because we shall often need to take the indefinite integral of different 
functions $F(y)$, we shall denote it with the symbol $I[F]$ as follows,
\begin{equation}
I[F](y) \equiv \int^y dy' F(y') \; .
\end{equation}
By consulting the expansions (\ref{Aexp}) and (\ref{gamexp}) we see that,
\begin{equation}
{\rm Res}\Bigl[y^{\frac{D}2-2} I[A' \gamma]\Bigr] = \frac{H^{D-2}}{4
\pi^{\frac{D}2}} \Gamma\Bigl(\frac{D}2 \!-\! 1\Bigr) \gamma(0) \; .
\end{equation}
Therefore the final term in (\ref{lastex}) gives precisely right residue 
contribution and we conclude that the first diagram of Fig.~1 can be
written as,
\begin{equation}
-4 e^2 \sqrt{-g} g^{\mu\rho} \sqrt{-g'} g^{\prime \nu \sigma}
i\Bigl[\mbox{}_{\rho} \Delta_{\sigma}\Bigr](x;x') \partial_{\mu} 
\partial'_{\nu} i\Delta(x;x') \!=\! 4 e^2 \sqrt{-g} \sqrt{-g'} \square 
I[A' \gamma] \, . \quad \label{left}
\end{equation}

The other diagrams of Fig.~1 are straightforward. The middle one is,
\begin{equation}
-i e^2 \sqrt{-g} g^{\mu\nu} i\Bigl[\mbox{}_{\mu} \Delta_{\nu}\Bigr](x;x')
\delta^D(x \!-\! x') = -i e^2 D \gamma(0) \sqrt{-g} \delta^D(x \!-\! x') \; .
\label{middle}
\end{equation}
The rightmost diagram of Fig.~1 represents field strength and conformal
renormalizations,
\begin{equation}
i \delta Z_2 \sqrt{-g} \square \delta^D(x \!-\! x') - i \delta \xi R
\sqrt{-g} \delta^D(x \!-\! x') \; . \label{right}
\end{equation}
Adding (\ref{left}), (\ref{middle}) and (\ref{right}) gives our result
for the $++$ polarization of the regulated one loop self-mass-squared,
\begin{eqnarray}
\lefteqn{-i M^2_{\scriptscriptstyle ++}(x;x') = 4 e^2 \sqrt{-g} \sqrt{-g'} 
\square I[A' \gamma] + i \delta Z_2 \sqrt{-g} \square \delta^D(x \!-\! x') }
\nonumber \\
& & \hspace{3.5cm} -i \Bigl[ D (D\!-\! 1) H^2 \delta \xi + e^2 D \gamma(0)\Bigr]
\sqrt{-g} \delta^D(x \!-\! x') \; . \qquad \label{M++reg}
\end{eqnarray}

To renormalize the one loop self-mass-squared we must isolate the divergences
implicit in the first term of (\ref{M++reg}). From the expansions (\ref{Aexp}) 
for $A'(y)$ and (\ref{gamnor}) for $\gamma(y)$ we compute,
\begin{eqnarray}
\lefteqn{A'(y) \gamma(y) = -\frac18 (D\!-\!1) \Gamma\Bigl(\frac{D}2\Bigr)
\Gamma\Bigl(\frac{D}2 \!-\! 1 \Bigr) \frac{H^{2 D-4}}{(4\pi)^{D}}
\Biggl\{ \Bigl(\frac{4}{y}\Bigr)^{D-1} \!\!\! + 3 \Bigl(\frac{4}{y}\Bigr)^2 + 3
\Bigl(\frac{4}{y}\Bigr) } \nonumber \\
& & \hspace{.5cm} + \Biggl[ \Bigl(\frac{4}{y}\Bigr)^2 + 4 \Bigl(\frac{4}{y}
\Bigr) + \frac{4}{1 \!-\! \frac{y}4} + \frac{3}{(1 \!-\! \frac{y}4)^2} \Biggr]
\ln\Bigl(\frac{y}4\Bigr) + \frac{3}{1 \!-\! \frac{y}4} + O\Bigl(\frac{D\!-\!4}{
y^2}\Bigr) \Biggr\} . \qquad
\end{eqnarray}
Hence the indefinite integral is,
\begin{eqnarray}
\lefteqn{I[A' \gamma](y) = \frac14 (D\!-\!1) \Gamma^2\Bigl(\frac{D}2 \!-\! 1
\Bigr) \frac{H^{2D-4}}{(4\pi)^{D}} \Biggl\{\Bigl(\frac{4}{y}\Bigr)^{D-2}
\!\!\! + 8 \Bigl(\frac{4}{y}\Bigr) + 8 \sum_{n=1}^{\infty} \frac1{n^2} 
\Bigl(\frac{y}4\Bigr)^n } \nonumber \\
& & \hspace{.5cm} + \Biggl[ 2 \Bigl(\frac{4}{y}\Bigr) - 4 \ln\Bigl(\frac{y}{4} 
\Bigr) + 8 \ln\Bigl(1 \!-\! \frac{y}4\Bigr) - \frac{6}{1 \!-\! \frac{y}4} 
\Biggr] \ln\Bigl(\frac{y}4\Bigr) + O\Bigl(\frac{D\!-\!4}{y}\Bigr) \Biggr\} . 
\qquad \label{IApg}
\end{eqnarray}

It is important to understand that one typically wants to use $-i 
M^2_{\scriptscriptstyle ++}(x;x')$ inside an integral over $x^{\prime \mu}$,
so an expression is ``renormalized'' when it has been written in a form which
is integrable (with respect to $x^{\prime \mu}$) in $D \!=\! 4$ spacetime
dimensions. Each term of (\ref{IApg}) except the first meets this requirement.
To isolate the divergence in this first term we exploit relation (\ref{dAlem}) 
to express it in terms of a lower power which integrable,
\begin{equation}
\Bigl(\frac{4}{y}\Bigr)^{D-2} = \frac2{(D\!-\!3)(D\!-\!4)} \frac{\square}{H^2} 
\Biggl\{\Bigl(\frac{4}{y}\Bigr)^{D-3} \Biggr\} - \frac{4}{D\!-\!4} 
\Bigl(\frac{4}{y}\Bigr)^{D-3} \; . \label{pint}
\end{equation}

The d`Alembertian with respect to $x^{\mu}$ can be pulled outside the 
integration over $x^{\prime \mu}$, so (\ref{pint}) really is an integrable 
expression. We could take the limit $D \rightarrow 4$ at this point, were 
it not for the factors of $1/(D\!-\!4)$. Of course these represent the
ultraviolet divergence we have been laboring to extract! To isolate it on
a local term suitable for renormalization, we add zero in the form,
\begin{equation}
0 = \frac{(4\pi)^{\frac{D}2} H^{-D}}{\Gamma(\frac{D}2 \!-\! 1)} 
\frac{i}{\sqrt{-g}} \delta^D(x \!-\! x') - \frac{\square}{H^2} \Biggl\{
\Bigl(\frac{4}{y}\Bigr)^{\frac{D}2-1} \Biggr\} + \frac{D}2 \Bigl(\frac{D}2
\!-\! 1\Bigr) \Bigl(\frac{4}{y}\Bigr)^{\frac{D}2-1} \; . \label{zero}
\end{equation}
Hence we conclude,
\begin{eqnarray}
\lefteqn{\Bigl(\frac{4}{y}\Bigr)^{D-2} \!\!\!= \frac{2}{(D \!-\! 3) (D\!-\!4)} 
\Biggl\{ \! \frac{(4\pi)^{\frac{D}2} H^{-D}}{\Gamma(\frac{D}2 \!-\! 1)} 
\frac{i \delta^D(x \!-\! x')}{\sqrt{-g}} \!+\!  \frac{\square}{H^2} \! \Biggl[
\! \Bigl(\frac{4}{y} \Bigr)^{D-3} \!\!\! - \Bigl(\frac{4}{y}\Bigr)^{\frac{D}2-
1} \! \Biggr] \! \Biggr\} } \nonumber \\
& & \hspace{5cm} - \frac{4}{D\!-\!4} \Biggl\{\Bigl(\frac{4}{y}\Bigr)^{D-3} 
\!\!\! - \frac{D (D\!-\! 2)}{8 (D\!-\!3)} \Bigl(\frac{4}{y}\Bigr)^{\frac{D}2-1} 
\Biggr\} \; , \qquad \\
& & = \frac{2}{(D \!-\! 3) (D\!-\!4)} \frac{(4\pi)^{\frac{D}2} H^{-D}}{
\Gamma(\frac{D}2 \!-\! 1)} \frac{i}{\sqrt{-g}} \delta^D(x \!-\! x') -
\frac{\square}{H^2} \Biggl\{ \frac{4}{y} \ln\Bigl(\frac{y}4\Bigr)\Biggr\}
\nonumber \\
& & \hspace{6cm} + 2 \Bigl(\frac{4}{y}\Bigr) \ln\Bigl(\frac{y}{4}\Bigr) -
\Bigl(\frac{4}{y} \Bigr) + O(D \!-\! 4) \; . \qquad \label{D-2}
\end{eqnarray}

Substituting (\ref{D-2}) in expression (\ref{IApg}), and the result into
(\ref{M++reg}), gives a form for the self-mass-squared from which 
renormalization parameters can be inferred,
\begin{eqnarray}
\lefteqn{-i M^2_{\scriptscriptstyle ++}(x;x') = i \Biggl\{ \delta Z_2 +
\frac{e^2 H^{D-4}}{(4 \pi)^4} \frac{2 (D\!-\!1) \Gamma(\frac{D}2 \!-\! 1)}{
(D\!-\!3)(D\!-\!4)} \Biggr\} \sqrt{-g} \square \delta^D(x \!-\! x') }
\nonumber \\
& &  -i \Bigl[ D (D\!-\! 1) H^2 \delta \xi + e^2 D \gamma(0)\Bigr]
\sqrt{-g} \delta^D(x \!-\! x') - \frac{3 e^2 H^2}{(4\pi)^4} \sqrt{-g} 
\sqrt{-g'} \nonumber \\
& & \hspace{2cm} \times \square^2 \Biggl\{ \frac{4}{y} \ln\Bigl(\frac{y}4\Bigr)
\Biggr\} + \frac{3 e^2 H^4}{(4 \pi)^4} \sqrt{-g} \sqrt{-g'} \square \Biggl\{ 7 
\Bigl(\frac{4}{y}\Bigr) + 8 \sum_{n=1}^{\infty} \frac1{n^2} \Bigl(\frac{y}4
\Bigr)^n \nonumber \\
& & \hspace{1cm} + \Biggl[ 4 \Bigl(\frac{4}{y}\Bigr) - 4 \ln\Bigl(\frac{y}{4} 
\Bigr) + 8 \ln\Bigl(1 \!-\! \frac{y}4\Bigr) - \frac{6}{1 \!-\! \frac{y}4} 
\Biggr] \ln\Bigl(\frac{y}4\Bigr) \Biggr\} + O(D\!-\!4) \; . \qquad 
\label{M++exp}
\end{eqnarray}
This fixes the one loop counterterms up to finite renormalizations,
\begin{eqnarray}
\delta \xi & = & -\frac{e^2 \gamma(0)}{(D\!-\!1) H^2} + 
\frac{12 e^2 \delta \xi_{\rm fin}}{D(D\!-\!1)} + O(e^4) \; , \label{delxi} \\
\delta Z_2 & = & -\frac{e^2 H^{D-4}}{(4\pi)^{\frac{D}2}} \frac{2 (D\!-\!1)
\Gamma(\frac{D}2 \!-\! 1)}{(D\!-\!3) (D\!-\!4)} + e^2 \delta Z_{\rm fin} 
+ O(e^4) \label{delZ2} \; .
\end{eqnarray}
Taking the limit $D \rightarrow 4$ gives the renormalized result at one loop
order,
\begin{eqnarray}
\lefteqn{-i M^2_{\scriptscriptstyle ++ \atop {\rm ren}}(x;x') = i e^2 \delta 
Z_{\rm fin} \sqrt{-g} \square \delta^4(x \!-\! x') -i 12 e^2 H^2 \delta 
\xi_{\rm fin} \sqrt{-g} \delta^4(x \!-\! x') } \nonumber \\
& & - \frac{3 e^2 H^2}{(4\pi)^4} \sqrt{-g} \sqrt{-g'} 
\square^2 \Biggl\{ \frac{4}{y} \ln\Bigl(\frac{y}4\Bigr)
\Biggr\} + \frac{3 e^2 H^4}{(4 \pi)^4} \sqrt{-g} \sqrt{-g'} \square \Biggl\{ 
7 \Bigl(\frac{4}{y}\Bigr) \nonumber \\
& & + 8 \sum_{n=1}^{\infty} \frac1{n^2} \Bigl(\frac{y}4 \Bigr)^n 
+ \Biggl[ 4 \Bigl(\frac{4}{y}\Bigr) - 4 \ln\Bigl(\frac{y}{4} 
\Bigr) + 8 \ln\Bigl(1 \!-\! \frac{y}4\Bigr) - \frac{6}{1 \!-\! \frac{y}4} 
\Biggr] \ln\Bigl(\frac{y}4\Bigr) \Biggr\} \; . \qquad \label{M++ren}
\end{eqnarray}

\section{$\langle \Omega \vert (D_{\mu} \varphi)^* D_{\nu} \varphi \vert 
\Omega \rangle$}

The purpose of this section is to evaluate the leftmost of the VEV's
in (\ref{twoVs}) at one and two loop orders. We begin by decomposing the
gauge invariant operator product into noninvariant components,
\begin{eqnarray}
(D_{\mu} \varphi)^* D_{\nu} \varphi & \equiv & \Bigl(\partial_{\mu} \!-\! 
i e A_{\mu} \Bigr) \varphi^* \Bigl(\partial_{\nu} \!+\! i e A_{\nu}\Bigr) 
\varphi \; , \\
& = & \partial_{\mu} \varphi^* \partial_{\nu} \varphi - i e \varphi^* 
A_{\mu} \partial_{\nu} \varphi + i e \partial_{\mu} \varphi^* A_{\nu}
\varphi + e^2 \varphi^* A_{\mu} A_{\nu} \varphi \; . \qquad \label{opex}
\end{eqnarray}
The next three sub-sections are devoted, respectively, to writing out
and partially evaluating the expectation values of the order $e^0$, $e^1$ 
and $e^2$ operators in (\ref{opex}). These results are combined in the final
sub-section, at which point an important cancellation occurs. We then give
explicit results for the remaining integrals.

\subsection{$\langle \Omega \vert \partial_{\mu} \varphi^* \partial_{\nu} 
\varphi \vert \Omega \rangle$}

The only diagram which contributes at one loop is Fig.~2.
\begin{center}
\begin{picture}(300,100)(0,0)
\DashArrowArc(150,50)(20,90,450){3}
\Vertex(150,70){3}
\Text(150,75)[b]{$x$}
\end{picture}
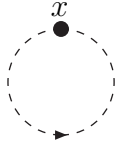
\\ {\rm Fig.~2: One loop contribution to $\langle \Omega \vert \partial_{\mu} 
\varphi^*(x) \partial_{\nu} \varphi(x) \vert \Omega \rangle$.}
\end{center}
It corresponds to the coincidence limit of the differentiated propagator,
whose value comes entirely from the $n=1$ term in (\ref{A}),
\begin{equation}
\lim_{x' \rightarrow x} \partial_{\mu} \partial'_{\nu} i\Delta(x;x') = 
\frac{H^{D-2}}{(4\pi)^{\frac{D}2}} \frac{\Gamma(D)}{\Gamma(\frac{D}2 \!+\! 1)}
\times -\frac12 H^2 g_{\mu\nu} = -\Bigl(\frac{D\!-\!1}{D}\Bigr) k H^2
g_{\mu\nu} \; . \label{1lpterm}
\end{equation}

Fig.~3 depicts the three diagrams which contribute at two loop order,
the scalar bilinear being an insertion at point $x^{\mu}$.
\begin{center}
\begin{picture}(390,100)(0,0)
\Photon(55,50)(125,50){3}{8}
\DashArrowArc(90,50)(35,0,90){3}
\DashArrowArc(90,50)(35,90,180){3}
\DashArrowArc(90,50)(35,180,360){3}
\Vertex(90,85){3}
\Text(90,93)[b]{$x$}
\Vertex(55,50){3}
\Text(42,52)[l]{$x'$}
\Vertex(125,50){3}
\Text(143,52)[r]{$x''$}
\Text(162.5,45)[b]{\Large $+$}
\DashArrowArc(210,50)(20,90,270){3}
\DashArrowArc(210,50)(20,-90,90){3}
\PhotonArc(210,20)(10,0,360){2}{8}
\Vertex(210,70){3}
\Text(210,75)[b]{$x$}
\Vertex(210,30){3}
\Text(210,35)[b]{$x'$}
\Text(257.5,45)[b]{\Large $+$}
\DashArrowArc(305,50)(20,90,270){3}
\DashArrowArc(305,50)(20,-90,90){3}
\Vertex(305,70){3}
\Text(305,75)[b]{$x$}
\Vertex(305,30){3}
\Text(306,23)[b]{\LARGE $\times$}
\Text(305,37)[b]{$x'$}
\end{picture}
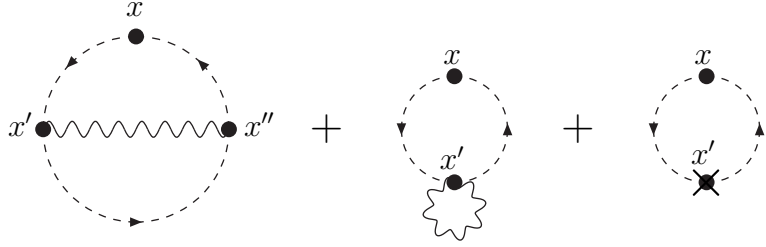
\\ {\rm Fig.~3: Two loop contributions to $\langle \Omega \vert \partial_{\mu} 
\varphi^*(x) \partial_{\nu} \varphi(x) \vert \Omega \rangle$.}
\end{center}
The alert reader will recognize these as the three diagrams of Fig.~2,
with the external lines joined. This means we can express the diagrams
in Fig.~3 as a sum over simple integrals involving the four one loop 
contributions to the self-mass-squared,
\begin{equation}
\sum_{\pm \pm} \int d^Dx' \partial_{\mu} i\Delta_{\scriptscriptstyle 
+\pm}(x;x') \int d^Dx'' \partial_{\nu} i\Delta_{\scriptscriptstyle +\pm}(x;x'')
\times -i M^2_{\scriptscriptstyle \pm\pm}(x';x'') \; . \label{Mform}
\end{equation}

For our purposes it is better not to use the renormalized self-mass-squared
(\ref{M++ren}), which is only valid inside integrals over suitably smooth
functions and in the limit $D \rightarrow 4$. It is instead superior to use
the exact, regulated result (\ref{M++reg}). Symmetrizing the d`Alembertian 
and taking account of our result (\ref{delxi}) for the conformal counterterm, 
we can write the $++$ polarity as,
\begin{eqnarray}
\lefteqn{-i M^2_{\scriptscriptstyle ++}(x';x'') = 2 e^2 \sqrt{-g'} \sqrt{-g''} 
\, (\square' \!+\! \square'') I[A' \gamma]\Bigl(y_{\scriptscriptstyle 
++}(x';x'')\Bigr) } \nonumber \\
& & \hspace{2cm} + i \delta Z_2 \sqrt{-g'} \square' \delta^D(x' \!-\! x'')
-i 12 e^2 H^2 \delta \xi_{\rm fin} \sqrt{-g'} \delta^D(x \!-\! x') \; . \qquad 
\label{M++}
\end{eqnarray}
The other polarizations give,
\begin{eqnarray}
\lefteqn{-i M^2_{\scriptscriptstyle +-}(x';x'') = -2 e^2 \sqrt{-g'} \sqrt{-g''} 
\, (\square' \!+\! \square'') I[A' \gamma]\Bigl(y_{\scriptscriptstyle 
+-}(x';x'')\Bigr) \; , } \\
\lefteqn{-i M^2_{\scriptscriptstyle -+}(x';x'') = -2 e^2 \sqrt{-g'} \sqrt{-g''} 
\, (\square' \!+\! \square'') I[A' \gamma]\Bigl(y_{\scriptscriptstyle 
-+}(x';x'')\Bigr) \; , } \\
\lefteqn{-i M^2_{\scriptscriptstyle --}(x';x'') = 2 e^2 \sqrt{-g'} \sqrt{-g''} 
\, (\square' \!+\! \square'') I[A' \gamma]\Bigl(y_{\scriptscriptstyle 
--}(x';x'')\Bigr) } \nonumber \\
& & \hspace{2cm} - i \delta Z_2 \sqrt{-g'} \square' \delta^D(x' \!-\! x'')
+i 12 e^2 H^2 \delta \xi_{\rm fin} \sqrt{-g'} \delta^D(x \!-\! x') \; . \qquad 
\label{M--}
\end{eqnarray}

The counterterms are simple on account of the delta functions. The contribution
to (\ref{Mform}) from the conformal counterterm is,
\begin{eqnarray}
\lefteqn{ -i 12 e^2 H^2 \delta \xi_{\rm fin} \int d^Dx' \sqrt{-g'} 
\Biggl\{\partial_{\mu} i\Delta_{\scriptscriptstyle ++}(x;x') \partial_{\nu} 
i\Delta_{\scriptscriptstyle ++}(x;x') } \nonumber \\
& & \hspace{6cm} - \partial_{\mu} i\Delta_{\scriptscriptstyle +-}(x;x') 
\partial_{\nu} i\Delta_{\scriptscriptstyle +-}(x;x')\Biggr\} \; . \qquad
\label{dxiterm}
\end{eqnarray}
We postpone evaluation of these integral to the end of this section. On the
other hand, the contribution to (\ref{Mform}) from the field strength 
counterterm can be evaluated right away,
\begin{eqnarray}
\lefteqn{ i \delta Z_2 \int d^Dx' \sqrt{-g'} \Biggl\{\partial_{\mu} i\Delta_{
\scriptscriptstyle ++}(x;x') \partial_{\nu} \square' i\Delta_{
\scriptscriptstyle ++}(x;x') } \nonumber \\
& & \hspace{6cm} - \partial_{\mu} i\Delta_{\scriptscriptstyle +-}(x;x') 
\partial_{\nu} \square' i\Delta_{\scriptscriptstyle +-}(x;x')\Biggr\} \qquad 
\nonumber \\
& & = - \delta Z_2 \lim_{x' \rightarrow x} \partial_{\mu} \partial'_{\nu}
i\Delta_{\scriptscriptstyle ++}(x;x') = \delta Z_2 \Bigl(\frac{D\!-\!1}{D}
\Bigr) k H^2 g_{\mu\nu} \; . \label{dZ2term}
\end{eqnarray}

The nonlocal contributions suggest a partial integration, which raises the
issue of surface terms. In the Schwinger-Keldysh formalism the various 
polarizations combine to completely cancel contributions from the future
time surface, and from the spatial surfaces \cite{TW7}. However, there are
nonzero contributions from the initial value surface. We shall assume that
order $e^2$ corrections to the state wave functional completely cancel these
surface contributions. Even if this is not correct, the surface contributions
should fall off exponentially \cite{OW2}.

Of course the point of partially integrating is to take advantage of the
simple results of acting d`Alembertians on the various propagators,
\begin{eqnarray}
\sqrt{-g'} \square' i\Delta_{\scriptscriptstyle ++}(x;x') = 
i \delta^D(x \!-\! x') & , & \sqrt{-g'} \square' i\Delta_{\scriptscriptstyle 
+-}(x;x') = 0 \; , \qquad \\
\sqrt{-g''} \square'' i\Delta_{\scriptscriptstyle ++}(x;x'') = 
i \delta^D(x \!-\! x'') & , & \sqrt{-g''} \square'' i\Delta_{\scriptscriptstyle
+-}(x;x'') = 0 \; . \qquad
\end{eqnarray}
One should also note that the derivative acting upon whichever outer leg
propagator contributes a delta function, will ultimately wind up acting
to undo the indefinite integral of $A' \gamma$,
\begin{eqnarray}
\lefteqn{\int \!\! d^Dx'' \!\sqrt{-g''} \, \partial_{\nu} \square'' i\Delta_{
\scriptscriptstyle ++}(x;x'') I[A'\gamma]\!\Bigl(y_{\scriptscriptstyle \pm +}
\!(x';x'')\!\Bigr) \!=\! \partial_{\nu} I[A'\gamma]\! \Bigl(y_{
\scriptscriptstyle \pm +}\!(x';x)\!\Bigr) \; , \qquad } \\
& & \hspace{5.7cm} = \frac{\partial y}{\partial x^{\nu}} A'\Bigl( y_{
\scriptscriptstyle +\pm}(x;x')\Bigr) \gamma\Bigl( y_{\scriptscriptstyle 
+\pm}(x;x')\Bigr) \; . \qquad
\end{eqnarray}
Hence the nonlocal terms give,
\begin{equation}
i 4 e^2 \int d^Dx' \sqrt{-g'} \frac{\partial y}{\partial x^{(\mu}} \Bigl\{
\partial_{\nu)} i\Delta_{\scriptscriptstyle ++}(x;x') A'(y_{\scriptscriptstyle
++}) \gamma(y_{\scriptscriptstyle ++}) - (+-)\Bigr\} \; . \label{NLterm}
\end{equation}
(Parenthesized indices are symmetrized throughout this paper.) We shall not 
evaluate this contribution because it cancels against one in the next 
sub-section.

\subsection{$-ie \langle \Omega \vert \varphi^*(x) A_{\mu}(x) \partial_{\nu} 
\varphi(x) \vert \Omega \rangle + ie \langle \Omega \vert \partial_{\mu} 
\varphi^*(x) A_{\nu}(x) \varphi(x) \vert \Omega \rangle$}

Both VEV's have the diagram topology shown in Fig.~3.
\begin{center}
\begin{picture}(300,100)(0,0)
\Photon(150,85)(150,15){3}{8}
\DashArrowArc(150,50)(35,-90,90){3}
\DashArrowArc(150,50)(35,90,270){3}
\Vertex(150,85){3}
\Text(150,93)[b]{$x$}
\Vertex(150,15){3}
\Text(150,0)[b]{$x'$}
\end{picture}
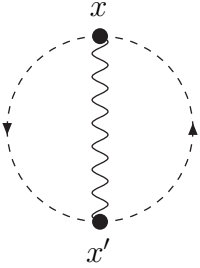
\\ {\rm Fig.~4: Lowest contribution to $-ie \langle \Omega \vert \varphi^*(x) 
A_{\mu}(x) \partial_{\nu} \varphi(x) \vert \Omega \rangle + ie \langle \Omega 
\vert \partial_{\mu} \varphi^*(x) A_{\nu}(x) \varphi(x) \vert \Omega \rangle$.}
\end{center}
The leftmost VEV can be simplified by another partial integration whose
surface term on the initial value surface we shall again assume is canceled 
by an order $e^2$ correction to the state wave functional,
\begin{eqnarray}
\lefteqn{i e^2 \int d^Dx' \sqrt{-g'} g^{\prime \rho\sigma} \Biggl\{
i\Bigl[\mbox{}_{\mu} \Delta_{\rho}\Bigr]_{\scriptscriptstyle ++}\!\!\!(x;x')
\Biggl[ \partial'_{\sigma} i\Delta_{\scriptscriptstyle ++}\!(x;x')
\partial_{\nu} i\Delta_{\scriptscriptstyle ++}\!(x;x') } \nonumber \\
& & \hspace{6cm} - i\Delta_{\scriptscriptstyle ++}\!(x;x') \partial_{\nu} 
\partial'_{\sigma} i\Delta_{\scriptscriptstyle ++}\!(x;x') \Biggr] - 
(+-)\Biggr\} \nonumber \\
& & \hspace{-.7cm} = i 2 e^2 \!\! \int \!\! d^Dx' \! \sqrt{-g'} g^{\prime \rho
\sigma} \Biggl\{\! i\Bigl[\mbox{}_{\mu} \Delta_{\rho}\Bigr]_{\scriptscriptstyle
++}\!\!\!(x;x') \partial'_{\sigma} i\Delta_{\scriptscriptstyle ++}\!(x;x')
\partial_{\nu} i\Delta_{\scriptscriptstyle ++}\!(x;x') \!-\! (+-) \!\Biggr\} . 
\qquad
\end{eqnarray}
The rightmost VEV makes the same contribution with the indices $\mu$ and
$\nu$ interchanged.

The next step is to act the $\partial'_{\sigma}$, carry out the
contraction using identities (\ref{ID1}-\ref{ID2}), and substitute 
(\ref{Brel}-\ref{Crel}). Because none of these operations depends upon the
Schwinger-Keldysh polarities we will carry them out generically,
\begin{eqnarray}
\lefteqn{g^{\prime \rho\sigma} i\Bigl[\mbox{}_{\mu} \Delta_{\rho}\Bigr]\!(x;x') 
\partial'_{\sigma} i\Delta(x;x') } \nonumber \\
& & \hspace{.7cm} = g^{\prime \rho\sigma} \Biggl\{ B(y) \frac{\partial^2 y}{
\partial x^{\mu} \partial x^{\prime \rho}} \!+\! C(y) \frac{\partial y}{
\partial x^{\mu}} \frac{\partial y}{\partial x^{\prime\rho}} \Biggr\} 
\Biggl\{ A'(y) \frac{\partial y}{\partial x^{\prime \sigma}} \!+\! k H a' 
\delta^0_{\sigma}\Biggr\} \; , \qquad \\
& & \hspace{.7cm} = H^2 \frac{\partial y}{\partial x^{\mu}} A'(y) \Bigl[ (2 
\!-\! y) B(y) \!+\! (4 y \!-\! y^2) C(y)\Bigr] - \frac{k H}{a'} i\Bigl[
\mbox{}_{\mu} \Delta_0 \Bigr]\!(x;x') \; , \qquad \\
& & \hspace{.7cm} = - \frac{\partial y}{\partial x^{\mu}} A'(y) \gamma(y)
- \frac{k H}{a'} i\Bigl[\mbox{}_{\mu} \Delta_0 \Bigr]\!(x;x') \; . \qquad
\end{eqnarray}
Therefore the two VEV's of this sub-section contribute,
\begin{eqnarray}
\lefteqn{-i 4 e^2 \int d^Dx' \sqrt{-g'} \frac{\partial y}{\partial x^{(\mu}} 
\Bigl\{\partial_{\nu)} i\Delta_{\scriptscriptstyle ++}(x;x') A'(y_{
\scriptscriptstyle ++}) \gamma(y_{\scriptscriptstyle ++}) - (+-)\Bigr\} }
\nonumber \\
& & \hspace{1.3cm} -i4 e^2k H \int d^Dx' a^{\prime D-1} \Bigl\{ \partial_{(\mu}
i \Delta_{\scriptscriptstyle ++}\!(x;x') i\Bigl[\mbox{}_{\nu)} \Delta_0 
\Bigr]\!(x;x') - (+-) \Bigr\} \; . \qquad \label{ordeterm}
\end{eqnarray} 
Of course the first line of (\ref{ordeterm}) just cancels (\ref{NLterm}), so
we need not evaluate either of them. We postpone evaluating the second term
until the end of this section.

\subsection{$e^2 \langle \Omega \vert \varphi^* A_{\mu} A_{\nu} \varphi \vert 
\Omega \rangle$}

The only order $e^2$ diagram that contributes to this VEV is depicted in
Fig.~5.
\begin{center}
\begin{picture}(300,100)(0,0)
\DashArrowArc(130,50)(20,0,360){3}
\PhotonArc(172,50)(20,-180,180){2}{15}
\Vertex(150,50){3}
\Text(140,50)[l]{$x$}
\end{picture}
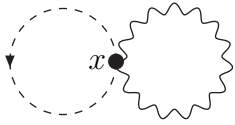
\\ {\rm Fig.~5: Lowest contribution to $e^2 \langle \Omega \vert \varphi^*(x) 
A_{\mu}(x) A_{\nu}(x) \varphi(x) \vert \Omega \rangle$.}
\end{center}
It is a straightforward coincidence limit,
\begin{equation}
e^2 i\Bigl[\mbox{}_{\mu} \Delta_{\nu}\Bigr](x;x) i\Delta(x;x) =
e^2 \gamma(0) \Bigl\{ A(0) + 2 k \ln(a) \Bigr\} g_{\mu\nu} \; . \label{esqterm}
\end{equation}

\subsection{The Final Result}

Adding the various contributions --- (\ref{1lpterm}), (\ref{dxiterm}),
(\ref{dZ2term}), (\ref{NLterm}), (\ref{ordeterm}) and (\ref{esqterm}) ---
gives the following result for the first VEV,
\begin{eqnarray}
\lefteqn{\Bigl\langle \Omega \Bigl\vert (D_{\mu} \varphi)^* D_{\nu} \varphi 
\Bigr\vert \Omega \Bigr\rangle \!=\! - \Bigl[1 \!-\! \delta Z_2\Bigr] 
\Bigl(\frac{D\!-\!1}{D}\Bigr) k H^2 g_{\mu\nu} \!+\! e^2 \gamma(0) \Bigl\{ A(0) 
\!+\! 2 k \ln(a) \Bigr\} g_{\mu\nu} } \nonumber \\
& & -i 12 e^2 H^2 \delta \xi_{\rm fin} \!\! \int \!\! d^Dx' \sqrt{-g'} 
\Bigl\{\partial_{\mu} i\Delta_{\scriptscriptstyle ++}\!(x;x') \partial_{\nu} 
i\Delta_{\scriptscriptstyle ++}\!(x;x') \!-\! (+-)\Bigr\} \nonumber \\
& & -i4 e^2k H \!\! \int \!\! d^Dx' a^{\prime D-1} \Bigl\{ \partial_{(\mu}
i \Delta_{\scriptscriptstyle ++}\!(x;x') i\Bigl[\mbox{}_{\nu)} \Delta_0 
\Bigr]_{\scriptscriptstyle ++}\!\!\!(x;x') \!-\! (+-) \Bigr\} + O(e^4) \; . 
\qquad \label{total}
\end{eqnarray}

It remains to evaluate the two final terms of (\ref{total}). For that purpose 
the 2nd covariant derivative of a function $f(y)$,
\begin{equation}
f_{;\mu\nu}(y) = H^2 (2 \!-\! y) f'(y) g_{\mu\nu} + f''(y) \frac{\partial y}{
\partial x^{\mu}} \frac{\partial y}{\partial x^{\nu}} \; ,
\end{equation}
implies an identity we shall employ many times,
\begin{equation}
F(y) \frac{\partial y}{\partial x^{\mu}} \frac{\partial y}{\partial x^{\nu}} 
= - H^2 g_{\mu\nu} (2 \!-\!y) I[F](y) + \Bigl[\partial_{\mu} \partial_{\nu}
\!-\! \Gamma^{\rho}_{~\mu\nu} \partial_{\rho}\Bigr] I^2[F](y) \; . \label{2ys}
\end{equation}
In using (\ref{2ys}), one extracts derivatives with respect to $x^{\mu}$ 
from integrations over $x^{\prime \mu}$. For example, consider the integral 
associated with the conformal counterterm in (\ref{total}),
\begin{eqnarray}
\lefteqn{\int \!\! d^Dx' \sqrt{-g'} \Bigl\{\partial_{\mu} i\Delta_{
\scriptscriptstyle ++}\!(x;x') \partial_{\nu} i\Delta_{\scriptscriptstyle ++}
\!(x;x') \!-\! (+-)\Bigr\} } \nonumber \\
& & \hspace{-.2cm} = \int \!\! d^Dx' \sqrt{-g'} \Biggl\{ \!\! \Bigl[A'(y_{
\scriptscriptstyle ++}) \frac{\partial y}{\partial x^{\mu}} \!+\! k H a 
\delta^0_{\mu}\Bigr] \! \Bigl[A'(y_{\scriptscriptstyle ++}) \frac{\partial y}{
\partial x^{\nu}} \!+\!  k H a \delta^0_{\nu}\Bigr] \!-\! (+-)\!\Biggr\} , 
\qquad \\
& & \hspace{-.2cm} = \int \!\! d^Dx' \sqrt{-g'} \Biggl\{ \frac{\partial y}{
\partial x^{\mu}} \frac{\partial y}{\partial x^{\nu}} A^{\prime 2}(y_{
\scriptscriptstyle ++}) \!+\! 2 k H a \delta^0_{(\mu} \frac{\partial y}{
\partial x^{\nu)}} A'(y_{\scriptscriptstyle ++}) - (+-)\Biggr\} , \qquad \\
& & \hspace{-.2cm} = -H^2 g_{\mu\nu} \int \!\! d^Dx' \sqrt{-g'} \Bigl[ (2 \!-\!
y_{\scriptscriptstyle ++}) I[A^{\prime 2}](y_{\scriptscriptstyle ++}) - 
(+-)\Bigr] \nonumber \\
& & \hspace{2cm} + \Bigl[\partial_{\mu} \partial_{\nu} \!-\! \Gamma^{\rho}_{
~\mu\nu} \partial_{\rho}\Bigr] \int \!\! d^Dx' \sqrt{-g'} \Bigl[ I^2[A^{\prime 
2}](y_{\scriptscriptstyle ++}) - (+-)\Bigr] \nonumber \\
& & \hspace{4cm} + 2 k H a \delta^0_{(\mu} \partial_{\nu)} \int \!\! d^Dx' 
\sqrt{-g'} \Bigl[ A(y_{\scriptscriptstyle ++}) - (+-)\Bigr] , \qquad \\
& & \hspace{-.2cm} = -H^2 g_{\mu\nu} \int \!\! d^Dx' \sqrt{-g'} \Bigl[ (2 \!-\!
y_{\scriptscriptstyle ++}) I[A^{\prime 2}](y_{\scriptscriptstyle ++}) - 
(+-)\Bigr] \nonumber \\
& & + H^2 \Bigl\{\! -g_{\mu\nu} \!+\! a^2 \delta^0_{\mu} \delta^0_{\nu}
[\partial_{\ln a} \!-\! 1] \Bigr\} \partial_{\ln a} \int \!\! d^Dx' \sqrt{-g'} 
\Bigl[ I^2[A^{\prime 2}](y_{\scriptscriptstyle ++}) - (+-)\Bigr] \nonumber \\
& & \hspace{3.5cm} + 2 k H^2 a^2 \delta^0_{\mu} \delta^0_{\nu} \, 
\partial_{\ln a} \int \!\! d^Dx' \sqrt{-g'} \Bigl[ A(y_{\scriptscriptstyle ++})
- (+-)\Bigr] . \qquad \label{dxiints}
\end{eqnarray}
Here $\partial_{\ln a}$ stands for the derivative with respect to $\ln(a)$,
which is the only coordinate upon which the integrals can depend.

Reducing the final term of (\ref{total}) to a similar form is facilitated
by two relations for temporal basis tensors,
\begin{equation}
\frac{\partial y}{\partial \eta'} = H a' (y \!-\! 2) + 2 H a \qquad {\rm and}
\qquad \frac{\partial^2 y}{\partial x^{\nu} \partial \eta'} = H a '
\frac{\partial y}{\partial x^{\nu}} + 2 H^2 a^2 \delta^0_{\nu} \; .
\end{equation}
Combining these with (\ref{genform}) and (\ref{Brel}-\ref{Crel}), we obtain
the relation,
\begin{eqnarray}
\lefteqn{i \Bigl[\mbox{}_{\nu} \Delta_0\Bigr](x;x') = B(y) \frac{\partial^2 y}{
\partial x^{\nu} \partial \eta'} + C \frac{\partial y}{\partial x^{\nu}}
\frac{\partial y}{\partial \eta'} \; , } \\
& & = H a' \frac{\partial y}{\partial x^{\nu}} \Bigl[ B(y) \!-\! (2 \!-\! y)
C(y)\Bigr] + 2 H a \frac{\partial y}{\partial x^{\nu}} C(y) + 2 H^2 a^2 
\delta^0_{\nu} B(y) \; , \qquad \\
& & = -\frac{a'}{(D\!-\!1) H} \frac{\partial y}{\partial x^{\nu}} \gamma'(y)
+ 2 H a \frac{\partial y}{\partial x^{\nu}} C(y) + 2 H^2 a^2 \delta^0_{\nu} 
B(y) \; . \qquad
\end{eqnarray}
Substituting this relation into the final term of (\ref{total}), decomposing 
the derivative of the scalar propagator, and making use of (\ref{2ys}) results 
in six integrals whose sum yields the final term of (\ref{total}),
\begin{eqnarray}
\lefteqn{I^1_{\mu\nu}\Bigl(\ln a\Bigr) \equiv \frac{i 4 e^2 k}{D \!-\! 1}
\int \!\! d^Dx' a^{\prime D} \frac{\partial y}{\partial x^{\mu}} \frac{\partial
y}{\partial x^{\nu}} \Bigl\{ A'(y_{\scriptscriptstyle ++}) \gamma'(y_{
\scriptscriptstyle ++}) - (+-)\Bigr\} \; , } \\
& & = -\frac{i 4 e^2 H^2 k}{D \!-\! 1} \Biggl\{ g_{\mu\nu} \!
\int \!\! d^Dx' a^{\prime D} \Bigl[ (2 \!-\! y_{\scriptscriptstyle ++}) I[A' 
\gamma'](y_{\scriptscriptstyle ++}) - (+-)\Bigr] \nonumber \\
& & \hspace{.5cm} + \Bigl[ g_{\mu\nu} \!-\! a^2 \delta^0_{\mu} \delta^0_{\nu} 
(\partial_{\ln a} \!-\! 1) \Bigr] \partial_{\ln a} \!\!\int \!\! d^Dx' 
a^{\prime D} \Bigl[ I^2[A' \gamma'](y_{\scriptscriptstyle ++}) - (+-)\Bigr] 
\Biggr\} , \qquad \label{I1} \\
\lefteqn{I^2_{\mu\nu}\Bigl(\ln a\Bigr) \equiv -i 8 e^2 k H^2 a \int \!\! d^Dx' 
a^{\prime D-1} \frac{\partial y}{\partial x^{\mu}} \frac{\partial y}{\partial 
x^{\nu}} \Bigl\{ A'(y_{\scriptscriptstyle ++}) C(y_{\scriptscriptstyle ++}) 
- (+-)\Bigr\} \; , } \\
& & = i 8 e^2 H^4 k a \Biggl\{ \! g_{\mu\nu} \!\! \int \!\! 
d^Dx' a^{\prime D-1} \! \Bigl[ (2 \!-\! y_{\scriptscriptstyle ++}) I[A' C]
(y_{\scriptscriptstyle ++}) \!-\! (+-)\Bigr] \nonumber \\
& & \hspace{.5cm} + \Bigl[ g_{\mu\nu} \!-\! a^2 \delta^0_{\mu} \delta^0_{\nu} 
(\partial_{\ln a} \!-\! 1) \Bigr] \partial_{\ln a} \!\! \int \!\! d^Dx' 
a^{\prime D-1} \Bigl[ I^2[A' C](y_{\scriptscriptstyle ++}) - (+-)\Bigr] 
\Biggr\} , \qquad \label{I2} \\
\lefteqn{I^3_{\mu\nu}\Bigl(\ln a\Bigr) \equiv - i 8 e^2 k H^3 a^2 \int \!\! 
d^Dx' a^{\prime D-1} \delta^0_{(\mu} \frac{\partial y}{\partial x^{\nu)}} 
\Bigl\{ A'(y_{\scriptscriptstyle ++}) B(y_{\scriptscriptstyle ++}) - (+-)
\Bigr\} \; , } \\
& & = -i 8 e^2 k H^4 a \, a^2 \delta^0_{\mu} \delta^0_{\nu} \, 
\partial_{\ln a} \! \int \!\! d^Dx' a^{\prime D-1} \Bigl\{ I[A' B](y_{
\scriptscriptstyle ++}) - (+-) \Bigr\} , \qquad \label{I3} \\
\lefteqn{I^4_{\mu\nu}\Bigl(\ln a\Bigr) \equiv \frac{i 4 e^2 k^2 H a}{D \!-\! 1}
\int \!\! d^Dx' a^{\prime D} \delta^0_{(\mu} \frac{\partial y}{\partial 
x^{\nu)}} \Bigl\{ \gamma'(y_{\scriptscriptstyle ++}) - (+-)\Bigr\} \; ,} \\
& & = \frac{i 4 e^2 k^2 H^2}{D \!-\! 1} \, a^2 \delta^0_{\mu} 
\delta^0_{\nu} \, \partial_{\ln a} \! \int \!\! d^Dx' a^{\prime D} \Bigl\{
\gamma(y_{\scriptscriptstyle ++}) - (+-)\Bigr\} \; , \\
\lefteqn{I^5_{\mu\nu}\Bigl(\ln a\Bigr) \equiv -i 8 e^2 k^2 H^3 a^2 \int \!\! 
d^Dx' a^{\prime D-1} \delta^0_{(\mu} \frac{\partial y}{\partial x^{\nu)}} 
\Bigl\{ C(y_{\scriptscriptstyle ++}) - (+-)\Bigr\} \; ,} \\
& & =- i 8 e^2 k^2 H^4 a \, a^2 \delta^0_{\mu} \delta^0_{\nu} \,
\partial_{\ln a} \! \int \!\! d^Dx' a^{\prime D-1} \Bigl\{ I[C](y_{
\scriptscriptstyle ++}) - (+-)\Bigr\} \; , \\
\lefteqn{I^6_{\mu\nu}\Bigl(\ln a\Bigr) \equiv -i 8 e^2 k^2 H^4 a \, a^2 
\delta^0_{\mu} \delta^0_{\nu} \, \int \!\! d^Dx' a^{\prime D-1} \Bigl\{ 
B(y_{\scriptscriptstyle ++}) - (+-)\Bigr\} \; .}
\end{eqnarray}

The next step is compute the various functionals of propagator functions
which appear in (\ref{dxiints}) and $I^{1-6}_{\mu\nu}$. Of course we
take $D\!=\!4$ on any terms which make finite contributions. For example,
the square of $A'(y)$ is,
\begin{equation}
A^{\prime 2}(y) = \frac{\Gamma^2(\frac{D}2)}{16} \frac{H^{2D-4}}{(4\pi)^D}
\Biggl\{ \Bigl(\frac{4}{y}\Bigr)^D + D \Bigl(\frac{4}{y}\Bigr)^{D-1}
+ 4 \Bigl(\frac{4}{y}\Bigr)^2 + O(D\!-\!4) \Biggr\} .
\end{equation}
Its integral is,
\begin{equation}
I[A^{\prime 2}](y) = \frac{\Gamma^2(\frac{D}2)}{4} \frac{H^{2D-4}}{(4\pi)^D}
\Biggl\{-\frac1{D\!-\!1} \Bigl(\frac{4}{y}\Bigr)^{D-1} \!\!- \frac{D}{D\!-\!2}
\Bigl(\frac{4}{y}\Bigr)^{D-2} \!\!- 4 \Bigl(\frac{4}{y}\Bigr) + O(D\!-\!4) 
\Biggr\} .
\end{equation}
Hence the integrands of the first two terms in (\ref{dxiints}) are,
\begin{eqnarray}
\lefteqn{(2 \!-\! y) I[A^{\prime 2}](y) = \frac{\Gamma^2(\frac{D}2)}{4} 
\frac{H^{2D-4}}{(4\pi)^D} \Biggl\{-\frac2{D\!-\!1}\Bigl(\frac{4}{y}\Bigr)^{D-1}
\!\!+ \Bigl[-\frac2{D\!-\!2} \!+\! \frac4{D\!-\!1}\Bigr] \Bigl(\frac{4}{y}
\Bigr)^{D-2} } \nonumber \\
& & \hspace{8.5cm} + 16 + O(D\!-\!4) \Biggr\} , \qquad \\
\lefteqn{I^2[A^{\prime 2}](y) = \Gamma^2\Bigl(\frac{D}2\Bigr) \frac{H^{2D-4}}{
(4 \pi)^D} \Biggl\{\frac1{(D\!-\!1) (D\!-\!2)} \Bigl(\frac{4}{y}\Bigr)^{D-2} }
\nonumber \\
& & \hspace{6cm} + 2 \Bigl(\frac{4}{y}\Bigr) - 4 \ln\Bigl(\frac{y}{4}\Bigr) 
+ O(D\!-\!4) \Biggr\} . \qquad
\end{eqnarray}
The integrand for the final term in (\ref{dxiints}) is just the $D\!=\!4$ 
limit of $A(y)$,
\begin{equation}
A(y) = \frac{H^2}{(4\pi)^2} \Biggl\{ \Bigl(\frac{4}{y}\Bigr) - 2 \ln\Bigl(
\frac{y}{4}\Bigr) - 1 + O(D\!-\!4)\Biggr\} .
\end{equation}

It turns out that each of the 11 integrands required for evaluating 
(\ref{dxiints}) and $I^{1-6}_{\mu\nu}$ can be expanded in terms of just
two potentially divergent functions of $y/4$ and eight finite ones. The
two potentially divergent functions are,
\begin{equation}
\Bigl(\frac{4}{y}\Bigr)^{D-1} \qquad {\rm and} \qquad 
\Bigl(\frac{4}{y}\Bigr)^{D-2} \; .
\end{equation}
In considering the finite functions it conserves space to set $x = y/4$.
These functions come from multiplying the less singular parts of propagators,
decomposing by partial fractions, and then performing whatever integrations
and subsequent multiplications are prescribed. The partial fractions 
decompositions are facilitated with the familiar identities,
\begin{eqnarray}
\frac1{x (1\!-\!x)} & = & \frac1{x} \!+\! \frac1{1\!-\!x} \; , \qquad \\
\frac1{x (1\!-\!x)^2} & = & \frac1{x} \!+\! \frac1{1\!-\!x} \!+\! 
\frac1{(1 \!-\! x)^2} \; , \qquad \\
\frac1{x (1\!-\!x)^3} & = & \frac1{x} \!+\! \frac1{1\!-\!x} \!+\! 
\frac1{(1 \!-\! x)^2} \!+\! \frac1{(1 \!-\! x)^3} \; , \qquad \\
\frac1{x^2 (1\!-\!x)} & = & \frac1{x^2} \!+\! \frac1{x} \!+\! \frac1{1 \!-\!x} 
\; , \qquad \\
\frac1{x^2 (1\!-\!x)^2} & = & \frac1{x^2} \!+\! \frac2{x} \!+\! \frac2{1\!-\!x} 
\!+\! \frac1{(1 \!-\! x)^2} \; , \qquad \\
\frac1{x^2 (1\!-\!x)^3} & = & \frac1{x^2} \!+\! \frac3{x} \!+\! 
\frac3{1 \!-\!x} \!+\! \frac2{(1 \!-\!x)^2} \!+\! \frac1{(1 \!-\!x)^3} 
\; . \qquad
\end{eqnarray}
The integrals we require are,
\begin{eqnarray}
\int dx \frac{\ln(x)}{x^2} & = & -\frac{\ln(x)}{x} \!-\! \frac1{x} 
\; , \qquad \\
\int dx \frac{\ln(x)}{x} & = & \frac12 \ln^2(x) \; , \qquad \\
\int dx \ln(x) & = & x \ln(x) \!-\! x \; , \qquad \\
\int dx \ln^2(x) & = & x \ln^2(x) \!-\! 2 x \ln(x) \!+\! 2 x \; , \qquad \\
\int dx \frac{\ln(x)}{(1 \!-\! x)^3} & = & \frac{\ln(x)}{2 (1\!-\!x)^2} \!-\!
\frac12 \ln(x) \!+\! \frac12 \ln(1 \!-\! x) - \frac1{2(1\!-\!x)} \; , \qquad \\
\int dx \frac{\ln(x)}{(1 \!-\! x)^2} & = & \frac{\ln(x)}{1\!-\!x} \!-\!
\ln(x) \!+\! \ln(1 \!-\! x) \; , \qquad \\
\int dx \frac{\ln(x)}{1 \!-\! x} & = & -\ln(1 \!-\! x) \ln(x) \!-\!
\sum_{n=1}^{\infty} \frac{x^n}{n^2} \; , \qquad \\
\int dx \ln(1 \!-\! x) \ln(x) & = & -(1 \!-\! x) \ln(1 \!-\! x) \ln(x) \!+\!
(1 \!-\! x) \ln(x) \nonumber \\
& & \hspace{5cm} + \sum_{n=1}^{\infty} \frac{x^{n+1}}{n (n\!+\!1)^2} 
\; . \qquad
\end{eqnarray}
Contributions which are analytic at $y \!=\! 0$ --- such as the infinite
sums --- cancel when we take the difference of $++$ and $+-$ terms.
So the eight finite functions can be taken to be,
\begin{eqnarray}
f_1(x) \equiv \frac1{x} & , & f_5(x) \equiv \ln^2(x) - 2 \ln(1\!-\!x) \ln(x) 
\; , \qquad \\
f_2(x) \equiv \frac{\ln(x)}{x} & , & f_6(x) \equiv x \ln^2(x) - 2 x \ln(1 
\!-\! x) \ln(x) - 2 \ln(x) \; , \qquad \\
f_3(x) \equiv \frac{\ln(x)}{(1 \!-\!x)^2} & , & f_7(x) \equiv \ln(x) \; , 
\qquad \\
f_4(x) \equiv \frac{\ln(x)}{1 \!-\! x} & , & f_8(x) \equiv \ln^2(x) \; . \qquad
\end{eqnarray}
We have tabulated the expansions for the integrands in (\ref{dxiints}), and 
those for integrals $I^{1-6}_{\mu\nu}$, in Tables~\ref{dexp} and \ref{fexp}. 
Table~\ref{dexp} gives the numerical coefficients of the potentially divergent 
contributions and Table~\ref{fexp} gives the finite contributions.

\begin{table}
\vbox{\tabskip=0pt \offinterlineskip
\def\tablerule{\noalign{\hrule}}
\halign to390pt {\strut#& \vrule#\tabskip=1em plus2em&
\hfil#\hfil& \vrule#& 
\hfil#\hfil& \vrule#& 
\hfil#\hfil& \vrule#& \hfil#\hfil&
\vrule#\tabskip=0pt\cr
\tablerule
\omit&height4pt&\omit&&\omit&&\omit&&\omit&\cr
&& $\!\!\!\!\! {\rm Integrand} \!\!\!\!\!$ && $\!\!\!\!\! {\rm Prefactor}
\!\!\!\!\!$ && $\!\!\!\!\! (\frac{4}{y})^{D-1} \!\!\!\!\!$ && $\!\!\!\!\! 
(\frac{4}{y})^{D-2} \!\!\!\!\!$ & \cr
\omit&height4pt&\omit&&\omit&&\omit&&\omit&\cr
\tablerule
\omit&height2pt&\omit&&\omit&&\omit&&\omit&\cr
\tablerule
\omit&height2pt&\omit&&\omit&&\omit&&\omit&\cr
&& $\!\!\!\!\! {\scriptstyle (2-y) I[A^{\prime 2}](y)} \!\!\!\!\!$ && 
$\!\!\!\!\! \frac{\Gamma^2(\frac{D}2) H^{2D-4}}{4 (4 \pi)^D} \!\!\!\!\!$ && 
$\!\!\!\!\! -\frac2{D-1} \!\!\!\!\!$ && $\!\!\!\!\! -\frac2{D-2} + 
\frac4{D-1} \!\!\!\!\!$ & \cr
\omit&height2pt&\omit&&\omit&&\omit&&\omit&\cr
\tablerule
\omit&height2pt&\omit&&\omit&&\omit&&\omit&\cr
&& $\!\!\!\!\! {\scriptstyle I^2[A^{\prime 2}](y)} \!\!\!\!\!$ && $\!\!\!\!\!
\frac{\Gamma^2(\frac{D}2) H^{2D-4}}{(4 \pi)^D} \!\!\!\!\!$ && $\!\!\!\!\! 0 
\!\!\!\!\!$ && $\!\!\!\!\! \frac1{(D-1)(D-2)} \!\!\!\!\!$ & \cr
\omit&height2pt&\omit&&\omit&&\omit&&\omit&\cr
\tablerule
\omit&height2pt&\omit&&\omit&&\omit&&\omit&\cr
&& $\!\!\!\!\! {\scriptstyle (2-y) I[A'\gamma'](y)} \!\!\!\!\!$ && $\!\!\!\!\!
\frac{(D-1) \Gamma^2(\frac{D}2) H^{2D-4}}{8 (4 \pi)^D} \!\!\!\!\!$ && 
$\!\!\!\!\! -\frac2{D-1} \!\!\!\!\!$ && $\!\!\!\!\! -2 + \frac4{D-1} + 
\frac4{(D-2)^2} \!\!\!\!\!$ & \cr
\omit&height2pt&\omit&&\omit&&\omit&&\omit&\cr
\tablerule
\omit&height2pt&\omit&&\omit&&\omit&&\omit&\cr
&& $\!\!\!\!\! {\scriptstyle I^2[A'\gamma'](y)} \!\!\!\!\!$ && $\!\!\!\!\!
\frac{(D-1) \Gamma^2(\frac{D}2) H^{2D-4}}{2 (4 \pi)^D} \!\!\!\!\!$ && 
$\!\!\!\!\! 0 \!\!\!\!\!$ && $\!\!\!\!\! \frac1{(D-1)(D-2)} \!\!\!\!\!$ & \cr
\omit&height2pt&\omit&&\omit&&\omit&&\omit&\cr
\tablerule
\omit&height2pt&\omit&&\omit&&\omit&&\omit&\cr
&& $\!\!\!\!\! {\scriptstyle (2-y) I[A'C](y)} \!\!\!\!\!$ && $\!\!\!\!\!
\frac{\Gamma^2(\frac{D}2) H^{2D-6}}{16 (4 \pi)^D} \!\!\!\!\!$ && 
$\!\!\!\!\! -\frac2{D-1} \!\!\!\!\!$ && $\!\!\!\!\! \frac4{D-1} 
-\frac{2D}{D-2} - \frac4{(D-2)^2} \!\!\!\!\!$ & \cr
\omit&height2pt&\omit&&\omit&&\omit&&\omit&\cr
\tablerule
\omit&height2pt&\omit&&\omit&&\omit&&\omit&\cr
&& $\!\!\!\!\! {\scriptstyle I^2[A'C](y)} \!\!\!\!\!$ && $\!\!\!\!\!
\frac{\Gamma^2(\frac{D}2) H^{2D-6}}{4 (4 \pi)^D} \!\!\!\!\!$ && 
$\!\!\!\!\! 0 \!\!\!\!\!$ && $\!\!\!\!\! \frac1{(D-1)(D-2)} \!\!\!\!\!$ & \cr
\omit&height2pt&\omit&&\omit&&\omit&&\omit&\cr
\tablerule
\omit&height2pt&\omit&&\omit&&\omit&&\omit&\cr
&& $\!\!\!\!\! {\scriptstyle I[A'B](y)} \!\!\!\!\!$ && $\!\!\!\!\!
\frac{\Gamma^2(\frac{D}2) H^{2D-6}}{4 (4 \pi)^D} \!\!\!\!\!$ 
&& $\!\!\!\!\! 0 \!\!\!\!\!$ && $\!\!\!\!\! -\frac2{(D-2)^2} \!\!\!\!\!$ & \cr
\omit&height2pt&\omit&&\omit&&\omit&&\omit&\cr
\tablerule
\omit&height2pt&\omit&&\omit&&\omit&&\omit&\cr
\tablerule}}

\caption{Potentially Divergent Contributions to Expansions of Integrands.}

\label{dexp}

\end{table}

The next step is to perform the integrations, against either $a^{\prime D}$
or $a^{\prime D-1}$, as the case may be. We have done this generically for
each of the two potentially divergent functions in Table~\ref{dint}. 
Table~\ref{f4int} gives the integrals of the eight finite functions times 
$a^{\prime 4}$. The analogous results for $a^{\prime 3}$ are reported in
Table~\ref{f3int}. The Appendix describes how these results were obtained.

\begin{table}
\vbox{\tabskip=0pt \offinterlineskip
\def\tablerule{\noalign{\hrule}}
\halign to390pt {\strut#& \vrule#\tabskip=1em plus2em&
\hfil#\hfil& \vrule#& 
\hfil#\hfil& \vrule#& 
\hfil#\hfil& \vrule#& 
\hfil#\hfil& \vrule#& 
\hfil#\hfil& \vrule#& 
\hfil#\hfil& \vrule#& 
\hfil#\hfil& \vrule#& 
\hfil#\hfil& \vrule#& \hfil#\hfil&
\vrule#\tabskip=0pt\cr
\tablerule
\omit&height4pt&\omit&&\omit&&\omit&&\omit&&\omit&&\omit&&\omit
&&\omit&&\omit&\cr
&& $\!\!\!\!\! {\rm Integrand} \!\!\!\!\!$ && $\!\!\!\!\! {\rm Prefactor}
\!\!\!\!\!$ && $\!\!\!\!\! f_1 \!\!\!\!\!$ && $\!\!\!\!\! f_2 \!\!\!\!\!$ 
&& $\!\!\!\!\! f_3 \!\!\!\!\!$ && $\!\!\!\!\! f_4 \!\!\!\!\!$ 
&& $\!\!\!\!\! f_5 \!\!\!\!\!$ && $\!\!\!\!\! f_6 \!\!\!\!\!$ 
&& $\!\!\!\!\! f_7 \!\!\!\!\!$ & \cr
\omit&height4pt&\omit&&\omit&&\omit&&\omit&&\omit&&\omit&&\omit
&&\omit&&\omit&\cr
\tablerule
\omit&height2pt&\omit&&\omit&&\omit&&\omit&&\omit&&\omit&&\omit
&&\omit&&\omit&\cr
\tablerule
\omit&height2pt&\omit&&\omit&&\omit&&\omit&&\omit&&\omit&&\omit
&&\omit&&\omit&\cr
&& $\!\!\!\!\! {\scriptstyle (2-y) I[A^{\prime 2}](y)} \!\!\!\!\!$ && 
$\!\!\!\!\! \frac{H^4}{4 (4 \pi)^4} \!\!\!\!\!$ && $\!\!\!\!\! 0 \!\!\!\!\!$ &&
$\!\!\!\!\! 0 \!\!\!\!\!$ && $\!\!\!\!\! 0 \!\!\!\!\!$ &&
$\!\!\!\!\! 0 \!\!\!\!\!$ && $\!\!\!\!\! 0 \!\!\!\!\!$ &&
$\!\!\!\!\! 0 \!\!\!\!\!$ && $\!\!\!\!\! 0 \!\!\!\!\!$ & \cr
\omit&height2pt&\omit&&\omit&&\omit&&\omit&&\omit&&\omit&&\omit
&&\omit&&\omit&\cr
\tablerule
\omit&height2pt&\omit&&\omit&&\omit&&\omit&&\omit&&\omit&&\omit
&&\omit&&\omit&\cr
&& $\!\!\!\!\! {\scriptstyle I^2[A^{\prime 2}](y)} \!\!\!\!\!$ && $\!\!\!\!\!
\frac{H^4}{(4 \pi)^4} \!\!\!\!\!$ && $\!\!\!\!\! 2 \!\!\!\!\!$ &&
$\!\!\!\!\! 0 \!\!\!\!\!$ && $\!\!\!\!\! 0 \!\!\!\!\!$ &&
$\!\!\!\!\! 0 \!\!\!\!\!$ && $\!\!\!\!\! 0 \!\!\!\!\!$ &&
$\!\!\!\!\! 0 \!\!\!\!\!$ && $\!\!\!\!\! -4 \!\!\!\!\!$ & \cr
\omit&height2pt&\omit&&\omit&&\omit&&\omit&&\omit&&\omit&&\omit&&
\omit&&\omit&\cr
\tablerule
\omit&height2pt&\omit&&\omit&&\omit&&\omit&&\omit&&\omit&&\omit&&
\omit&&\omit&\cr
&& $\!\!\!\!\! {\scriptstyle A(y)} \!\!\!\!\!$ && $\!\!\!\!\!
\frac{H^2}{(4 \pi)^2} \!\!\!\!\!$ && $\!\!\!\!\! 1 \!\!\!\!\!$ &&
$\!\!\!\!\! 0 \!\!\!\!\!$ && $\!\!\!\!\! 0 \!\!\!\!\!$ &&
$\!\!\!\!\! 0 \!\!\!\!\!$ && $\!\!\!\!\! 0 \!\!\!\!\!$ &&
$\!\!\!\!\! 0 \!\!\!\!\!$ && $\!\!\!\!\! -2 \!\!\!\!\!$ & \cr
\omit&height2pt&\omit&&\omit&&\omit&&\omit&&\omit&&\omit&&\omit&&
\omit&&\omit&\cr
\tablerule
\omit&height2pt&\omit&&\omit&&\omit&&\omit&&\omit&&\omit&&\omit&&
\omit&&\omit&\cr
&& $\!\!\!\!\! {\scriptstyle (2-y) I[A'\gamma'](y)} \!\!\!\!\!$ && $\!\!\!\!\!
\frac{3 H^4}{8 (4 \pi)^4} \!\!\!\!\!$ && $\!\!\!\!\! 13 \!\!\!\!\!$ &&
$\!\!\!\!\! 4 \!\!\!\!\!$ && $\!\!\!\!\! 6 \!\!\!\!\!$ &&
$\!\!\!\!\! 4 \!\!\!\!\!$ && $\!\!\!\!\! -10 \!\!\!\!\!$ &&
$\!\!\!\!\! 20 \!\!\!\!\!$ && $\!\!\!\!\! 0 \!\!\!\!\!$ & \cr
\omit&height2pt&\omit&&\omit&&\omit&&\omit&&\omit&&\omit&&\omit&&
\omit&&\omit&\cr
\tablerule
\omit&height2pt&\omit&&\omit&&\omit&&\omit&&\omit&&\omit&&\omit&&
\omit&&\omit&\cr
&& $\!\!\!\!\! {\scriptstyle I^2[A'\gamma'](y)} \!\!\!\!\!$ && $\!\!\!\!\!
\frac{3 H^4}{2 (4 \pi)^4} \!\!\!\!\!$ && $\!\!\!\!\! \frac12 \!\!\!\!\!$ &&
$\!\!\!\!\! 0 \!\!\!\!\!$ && $\!\!\!\!\! 0 \!\!\!\!\!$ &&
$\!\!\!\!\! -3 \!\!\!\!\!$ && $\!\!\!\!\! 1 \!\!\!\!\!$ &&
$\!\!\!\!\! -5 \!\!\!\!\!$ && $\!\!\!\!\! 0 \!\!\!\!\!$ & \cr
\omit&height2pt&\omit&&\omit&&\omit&&\omit&&\omit&&\omit&&\omit&&
\omit&&\omit&\cr
\tablerule
\omit&height2pt&\omit&&\omit&&\omit&&\omit&&\omit&&\omit&&\omit&&
\omit&&\omit&\cr
&& $\!\!\!\!\! {\scriptstyle (2-y) I[A'C](y)} \!\!\!\!\!$ && $\!\!\!\!\!
\frac{H^2}{16 (4 \pi)^4} \!\!\!\!\!$ && $\!\!\!\!\! -20 \!\!\!\!\!$ &&
$\!\!\!\!\! -8 \!\!\!\!\!$ && $\!\!\!\!\! -6 \!\!\!\!\!$ &&
$\!\!\!\!\! -16 \!\!\!\!\!$ && $\!\!\!\!\! 18 \!\!\!\!\!$ &&
$\!\!\!\!\! -36 \!\!\!\!\!$ && $\!\!\!\!\! 0 \!\!\!\!\!$ & \cr
\omit&height2pt&\omit&&\omit&&\omit&&\omit&&\omit&&\omit&&\omit&&
\omit&&\omit&\cr
\tablerule
\omit&height2pt&\omit&&\omit&&\omit&&\omit&&\omit&&\omit&&\omit&&
\omit&&\omit&\cr
&& $\!\!\!\!\! {\scriptstyle I^2[A'C](y)} \!\!\!\!\!$ && $\!\!\!\!\!
\frac{H^2}{4 (4 \pi)^4} \!\!\!\!\!$ && $\!\!\!\!\! \frac52 \!\!\!\!\!$ &&
$\!\!\!\!\! 0 \!\!\!\!\!$ && $\!\!\!\!\! 0 \!\!\!\!\!$ &&
$\!\!\!\!\! 3 \!\!\!\!\!$ && $\!\!\!\!\! -2 \!\!\!\!\!$ &&
$\!\!\!\!\! 9 \!\!\!\!\!$ && $\!\!\!\!\! 0 \!\!\!\!\!$ & \cr
\omit&height2pt&\omit&&\omit&&\omit&&\omit&&\omit&&\omit&&\omit&&
\omit&&\omit&\cr
\tablerule
\omit&height2pt&\omit&&\omit&&\omit&&\omit&&\omit&&\omit&&\omit&&
\omit&&\omit&\cr
&& $\!\!\!\!\! {\scriptstyle I[A'B](y)} \!\!\!\!\!$ && $\!\!\!\!\!
\frac{H^2}{4 (4 \pi)^4} \!\!\!\!\!$ && $\!\!\!\!\! -6 \!\!\!\!\!$ &&
$\!\!\!\!\! -3 \!\!\!\!\!$ && $\!\!\!\!\! 0 \!\!\!\!\!$ &&
$\!\!\!\!\! 3 \!\!\!\!\!$ && $\!\!\!\!\! 5 \!\!\!\!\!$ &&
$\!\!\!\!\! 0 \!\!\!\!\!$ && $\!\!\!\!\! 0 \!\!\!\!\!$ & \cr
\omit&height2pt&\omit&&\omit&&\omit&&\omit&&\omit&&\omit&&\omit&&
\omit&&\omit&\cr
\tablerule
\omit&height2pt&\omit&&\omit&&\omit&&\omit&&\omit&&\omit&&\omit&&
\omit&&\omit&\cr
&& $\!\!\!\!\! {\scriptstyle \gamma(y)} \!\!\!\!\!$ && $\!\!\!\!\!
\frac{3 H^2}{2 (4 \pi)^2} \!\!\!\!\!$ && $\!\!\!\!\! 1 \!\!\!\!\!$ &&
$\!\!\!\!\! 0 \!\!\!\!\!$ && $\!\!\!\!\! 1 \!\!\!\!\!$ &&
$\!\!\!\!\! 0 \!\!\!\!\!$ && $\!\!\!\!\! 0 \!\!\!\!\!$ &&
$\!\!\!\!\! 0 \!\!\!\!\!$ && $\!\!\!\!\! 0 \!\!\!\!\!$ & \cr
\omit&height2pt&\omit&&\omit&&\omit&&\omit&&\omit&&\omit&&\omit&&
\omit&&\omit&\cr
\tablerule
\omit&height2pt&\omit&&\omit&&\omit&&\omit&&\omit&&\omit&&\omit&&
\omit&&\omit&\cr
&& $\!\!\!\!\! {\scriptstyle I[C](y)} \!\!\!\!\!$ && $\!\!\!\!\!
\frac1{4 (4 \pi)^2} \!\!\!\!\!$ && $\!\!\!\!\! 1 \!\!\!\!\!$ &&
$\!\!\!\!\! 0 \!\!\!\!\!$ && $\!\!\!\!\! -1 \!\!\!\!\!$ &&
$\!\!\!\!\! -2 \!\!\!\!\!$ && $\!\!\!\!\! 0 \!\!\!\!\!$ &&
$\!\!\!\!\! 0 \!\!\!\!\!$ && $\!\!\!\!\! 0 \!\!\!\!\!$ & \cr
\omit&height2pt&\omit&&\omit&&\omit&&\omit&&\omit&&\omit&&\omit&&
\omit&&\omit&\cr
\tablerule
\omit&height2pt&\omit&&\omit&&\omit&&\omit&&\omit&&\omit&&\omit&&
\omit&&\omit&\cr
&& $\!\!\!\!\! {\scriptstyle B(y)} \!\!\!\!\!$ && $\!\!\!\!\!
\frac1{4 (4 \pi)^2} \!\!\!\!\!$ && $\!\!\!\!\! -1 \!\!\!\!\!$ &&
$\!\!\!\!\! 0 \!\!\!\!\!$ && $\!\!\!\!\! -1 \!\!\!\!\!$ &&
$\!\!\!\!\! -2 \!\!\!\!\!$ && $\!\!\!\!\! 0 \!\!\!\!\!$ &&
$\!\!\!\!\! 0 \!\!\!\!\!$ && $\!\!\!\!\! 0 \!\!\!\!\!$ & \cr
\omit&height2pt&\omit&&\omit&&\omit&&\omit&&\omit&&\omit&&\omit&&
\omit&&\omit&\cr
\tablerule
\omit&height2pt&\omit&&\omit&&\omit&&\omit&&\omit&&\omit&&\omit&&
\omit&&\omit&\cr
\tablerule}}

\caption{Finite Contributions to Expansions of Integrands.}

\label{fexp}

\end{table}

\begin{table}
\vbox{\tabskip=0pt \offinterlineskip
\def\tablerule{\noalign{\hrule}}
\halign to390pt {\strut#& \vrule#\tabskip=1em plus2em&
\hfil#\hfil& \vrule#& \hfil#\hfil& \vrule#& \hfil#\hfil&
\vrule#\tabskip=0pt\cr
\tablerule
\omit&height4pt&\omit&&\omit&&\omit&\cr
\omit&height2pt&\omit&&\omit&&\omit&\cr
&& $\!\!\!\!\! f(x) \!\!\!\!\!$ && $\!\!\!\!\! \int d^Dx' a^{\prime D} 
\{f(\frac{y_{++}}{4}) - f(\frac{y_{+-}}{4})\} \!\!\!\!\!$ && $\!\!\!\!\! 
\int d^Dx' a^{\prime D-1} \{f(\frac{y_{++}}{4}) - f(\frac{y_{+-}}{4})\} 
\!\!\!\!\!$ & \cr
\omit&height4pt&\omit&&\omit&&\omit&\cr
\tablerule
\omit&height2pt&\omit&&\omit&&\omit&\cr
\tablerule
\omit&height2pt&\omit&&\omit&&\omit&\cr
&& $\!\!\!\!\! (\frac1{x})^{D-1} \!\!\!\!\!$ && $\!\!\!\!\!
- \frac{4}{(D-2)(D-4)} + O(\frac1{a},{\scriptstyle D-4}) \!\!\!\!\!$ 
&& $\!\!\!\!\! 0 + O(\frac1{a^2},{\scriptstyle D-4}) \!\!\!\!\!$ & \cr
\omit&height2pt&\omit&&\omit&&\omit&\cr
\tablerule
\omit&height2pt&\omit&&\omit&&\omit&\cr
&& $\!\!\!\!\! (\frac1{x})^{D-2} \!\!\!\!\!$ && $\!\!\!\!\!
\frac{2}{D-4} + O(\frac1{a},{\scriptstyle D-4}) \!\!\!\!\!$ 
&& $\!\!\!\!\! \frac{2 a^{-1}}{(D-3)(D-4)} + O(\frac1{a^2},
{\scriptstyle D-4}) \!\!\!\!\!$ & \cr
\omit&height2pt&\omit&&\omit&&\omit&\cr
\tablerule
\omit&height2pt&\omit&&\omit&&\omit&\cr
\tablerule}}

\caption{Divergent Integrals. Multiply each term by $\frac{i (4\pi)^{\frac{D}2}
}{H^D \Gamma(\frac{D}2-1)}$.}

\label{dint}

\end{table}

\begin{table}
\vbox{\tabskip=0pt \offinterlineskip
\def\tablerule{\noalign{\hrule}}
\halign to390pt {\strut#& \vrule#\tabskip=1em plus2em&
\hfil#\hfil& \vrule#& \hfil#\hfil& \vrule#\tabskip=0pt\cr
\tablerule
\omit&height4pt&\omit&&\omit&\cr
\omit&height2pt&\omit&&\omit&\cr
&& $\!\!\!\!\! f(x) \!\!\!\!\!$ && $\!\!\!\!\! \int d^4x' a^{\prime 4} 
\{f(\frac{y_{++}}{4}) - f(\frac{y_{+-}}{4})\} \!\!\!\!\!$ & \cr
\omit&height4pt&\omit&&\omit&\cr
\tablerule
\omit&height2pt&\omit&&\omit&\cr
\tablerule
\omit&height2pt&\omit&&\omit&\cr
&& $\!\!\!\!\! \frac1{x} \!\!\!\!\!$ && $\!\!\!\!\!
- \frac12 + O(\frac1{a}) \!\!\!\!\!$ & \cr
\omit&height2pt&\omit&&\omit&\cr
\tablerule
\omit&height2pt&\omit&&\omit&\cr
&& $\!\!\!\!\! \frac{\ln(x)}{x} \!\!\!\!\!$ && $\!\!\!\!\!
\frac34 + O(\frac1{a}) \!\!\!\!\!$ & \cr
\omit&height2pt&\omit&&\omit&\cr
\tablerule
\omit&height2pt&\omit&&\omit&\cr
&& $\!\!\!\!\! \frac{\ln(x)}{(1-x)^2} \!\!\!\!\!$ && $\!\!\!\!\!
-\frac32 + \frac{\pi^2}6 + O(\frac1{a}) \!\!\!\!\!$ & \cr
\omit&height2pt&\omit&&\omit&\cr
\tablerule
\omit&height2pt&\omit&&\omit&\cr
&& $\!\!\!\!\! \frac{\ln(x)}{1-x} \!\!\!\!\!$ && $\!\!\!\!\!
\frac14 + O(\frac{\ln(a)}{a}) \!\!\!\!\!$ & \cr
\omit&height2pt&\omit&&\omit&\cr
\tablerule
\omit&height2pt&\omit&&\omit&\cr
&& $\!\!\!\!\! {\scriptstyle \ln^2(x) - 2 \ln(1-x) \ln(x)} \!\!\!\!\!$ && 
$\!\!\!\!\! \frac16 - \frac{\pi^2}9 + O(\frac{\ln(a)}{a}) \!\!\!\!\!$ & \cr
\omit&height2pt&\omit&&\omit&\cr
\tablerule
\omit&height2pt&\omit&&\omit&\cr
&& $\!\!\!\!\! {\scriptstyle  x \ln^2(x) - 2 x \ln(1-x) \ln(x) - 2 \ln(x)} 
\!\!\!\!\!$ && $\!\!\!\!\! \frac5{24} - \frac{\pi^2}{18} + O(\frac{\ln(a)}{a}) 
\!\!\!\!\!$ & \cr
\omit&height2pt&\omit&&\omit&\cr
\tablerule
\omit&height2pt&\omit&&\omit&\cr
&& $\!\!\!\!\! {\scriptstyle  \ln(x)} \!\!\!\!\!$ && $\!\!\!\!\! \frac16 
{\scriptstyle \ln(a)} - \frac{11}{36} + O(\frac1{a}) \!\!\!\!\!$ & \cr
\omit&height2pt&\omit&&\omit&\cr
\tablerule
\omit&height2pt&\omit&&\omit&\cr
&& $\!\!\!\!\! {\scriptstyle  \ln^2(x)} \!\!\!\!\!$ && $\!\!\!\!\! \frac16 
{\scriptstyle \ln^2(a)} - \frac89 {\scriptstyle \ln(a)} + \frac74 - 
\frac{\pi^2}{9} + O(\frac{\ln(a)}{a}) \!\!\!\!\!$ & \cr
\omit&height2pt&\omit&&\omit&\cr
\tablerule
\omit&height2pt&\omit&&\omit&\cr
\tablerule}}

\caption{Finite de Sitter Invariant Integrals. Multiply each term by 
$\frac{i 16 \pi^2}{H^4}$.}

\label{f4int}

\end{table}

\begin{table}
\vbox{\tabskip=0pt \offinterlineskip
\def\tablerule{\noalign{\hrule}}
\halign to390pt {\strut#& \vrule#\tabskip=1em plus2em&
\hfil#\hfil& \vrule#& \hfil#\hfil& \vrule#\tabskip=0pt\cr
\tablerule
\omit&height4pt&\omit&&\omit&\cr
\omit&height2pt&\omit&&\omit&\cr
&& $\!\!\!\!\! f(x) \!\!\!\!\!$ && $\!\!\!\!\! \int d^4x' a^{\prime 3}
\{f(\frac{y_{++}}{4}) - f(\frac{y_{+-}}{4})\} \!\!\!\!\!$ & \cr
\omit&height4pt&\omit&&\omit&\cr
\tablerule
\omit&height2pt&\omit&&\omit&\cr
\tablerule
\omit&height2pt&\omit&&\omit&\cr
&& $\!\!\!\!\! \frac1{x} \!\!\!\!\!$ && $\!\!\!\!\! -{\scriptstyle \ln(a)}
+ {\scriptstyle 1} + O(\frac1{a}) \!\!\!\!\!$ & \cr
\omit&height2pt&\omit&&\omit&\cr
\tablerule
\omit&height2pt&\omit&&\omit&\cr
&& $\!\!\!\!\! \frac{\ln(x)}{x} \!\!\!\!\!$ && $\!\!\!\!\! -\frac12
{\scriptstyle \ln^2(a)} + {\scriptstyle 2 \ln(a)} - {\scriptstyle 3} + 
\frac{\pi^2}3 + O(\frac{\ln(a)}{a}) \!\!\!\!\!$ & \cr
\omit&height2pt&\omit&&\omit&\cr
\tablerule
\omit&height2pt&\omit&&\omit&\cr
&& $\!\!\!\!\! \frac{\ln(x)}{(1-x)^2} \!\!\!\!\!$ && $\!\!\!\!\!
{\scriptstyle \ln(a)} - {\scriptstyle 1} - \frac{\pi^2}{6}
+ O(\frac1{a}) \!\!\!\!\!$ & \cr
\omit&height2pt&\omit&&\omit&\cr
\tablerule
\omit&height2pt&\omit&&\omit&\cr
&& $\!\!\!\!\! \frac{\ln(x)}{1-x} \!\!\!\!\!$ && $\!\!\!\!\!
\frac12 {\scriptstyle \ln^2(a)} - {\scriptstyle 2 \ln(a)} +
{\scriptstyle 3} + O(\frac{\ln(a)}{a}) \!\!\!\!\!$ & \cr
\omit&height2pt&\omit&&\omit&\cr
\tablerule
\omit&height2pt&\omit&&\omit&\cr
&& $\!\!\!\!\! {\scriptstyle \ln^2(x) - 2 \ln(1-x) \ln(x)} \!\!\!\!\!$ && 
$\!\!\!\!\! -{\scriptstyle \ln^2(a)} + {\scriptstyle 2 \ln(a)} - 
{\scriptstyle 5} + \frac{\pi^2}3 + O(\frac1{a}) \!\!\!\!\!$ & \cr
\omit&height2pt&\omit&&\omit&\cr
\tablerule
\omit&height2pt&\omit&&\omit&\cr
&& $\!\!\!\!\! {\scriptstyle  x \ln^2(x) - 2 x \ln(1-x) \ln(x) - 2 \ln(x)} 
\!\!\!\!\!$ && $\!\!\!\!\! -\frac12 {\scriptstyle \ln^2(a)} + \frac32
{\scriptstyle \ln(a)} - \frac{35}{12} + \frac{\pi^2}9 + O(\frac{\ln(a)}{a}) 
\!\!\!\!\!$ & \cr
\omit&height2pt&\omit&&\omit&\cr
\tablerule
\omit&height2pt&\omit&&\omit&\cr
&& $\!\!\!\!\! {\scriptstyle  \ln(x)} \!\!\!\!\!$ && $\!\!\!\!\! \frac16 
{\scriptstyle a} - \frac12 {\scriptstyle \ln(a)} + \frac14 + O(\frac1{a}) 
\!\!\!\!\!$ & \cr
\omit&height2pt&\omit&&\omit&\cr
\tablerule
\omit&height2pt&\omit&&\omit&\cr
&& $\!\!\!\!\! {\scriptstyle  \ln^2(x)} \!\!\!\!\!$ && $\!\!\!\!\! \frac13
{\scriptstyle a \ln(a)} \!-\! \frac{11}9 {\scriptstyle a} \!-\! \frac12 
{\scriptstyle \ln^2(a)} \!+\! {\scriptstyle 2 \ln(a)} \!-\! \frac94
\!+\! \frac{\pi^2}{3} \!+\! O(\frac{\ln(a)}{a}) \!\!\!\!\!$ & \cr
\omit&height2pt&\omit&&\omit&\cr
\tablerule
\omit&height2pt&\omit&&\omit&\cr
\tablerule}}

\caption{Finite de Sitter Breaking Integrals. Multiply each term by 
$\frac{i 16 \pi^2}{H^4 a}$.}

\label{f3int}

\end{table}

One evaluates each of the 11 integrals in (\ref{dxiints}) and $I^{1-6}_{
\mu\nu}$ by multiplying the appropriate tabulated results, and then acting 
whatever derivatives are prescribed. For example, the first term in 
(\ref{dxiints}) gives,
\begin{eqnarray}
\lefteqn{-H^2 g_{\mu\nu} \times \frac{\Gamma^2(\frac{D}2)}{4} \frac{H^{2D-4}}{
(4\pi)^{D}} \times \frac{i (4 \pi)^{\frac{D}2}}{ H^D \Gamma(\frac{D}2 \!-\!1)}
\Biggl\{ -\frac2{D\!-\!1} \times \frac{-4}{(D\!-\!2)(D\!-\!4)} } \nonumber \\
& & \hspace{2cm} + \Bigl[-\frac2{D\!-\!2} \!+\! \frac4{D\!-\!1}\Bigr] \times
\frac2{D \!-\! 4} \Biggr\} = -\frac{ i H^{D-2}}{(4 \pi)^{\frac{D}2}}
\frac{\Gamma(\frac{D}2)}{2 (D\!-\!4)} \, g_{\mu\nu} \; . \qquad 
\end{eqnarray}
The second term of (\ref{dxiints}) gives an even simpler result,
\begin{eqnarray}
\lefteqn{H^2 \Bigl\{\! -g_{\mu\nu} \!+\! a^2 \delta^0_{\mu} \delta^0_{\nu} 
[\partial_{\ln a} \!-\! 1] \Bigr\} \partial_{\ln a} \times 
\Gamma^2\Bigl(\frac{D}2\Bigr) \frac{H^{2D-4}}{(4\pi)^{D}} \times 
\frac{i (4 \pi)^{\frac{D}2}}{ H^D \Gamma(\frac{D}2\!-\!1)} } \nonumber \\
& & \hspace{2cm} \times \Biggl\{ \frac1{(D\!-\!1)(D\!-\!2)} \times 
\frac2{D\!-\!4} + 2 \times -\frac12 - 4 \times \Bigl[ \frac16 \ln(a) - 
\frac{11}{36}\Bigr] \Biggr\} \nonumber \\
& & = \frac{i H^2}{24 \pi^2} \Bigl[ g_{\mu\nu} + a^2 \delta^0_{\mu}
\delta^0_{\nu}\Bigr] . \qquad 
\end{eqnarray}
The final term of (\ref{dxiints}) is finite from the start,
\begin{equation}
2 k H^2 a^2 \delta^0_{\mu} \delta^0_{\nu} \, \partial_{\ln a} \times
\frac{H^2}{16 \pi^2} \times \frac{i 16 \pi^2}{H^4} \Biggl\{-\frac12
- 2 \times \Bigl[\frac16 \ln(a) - \frac{11}{36}\Bigr] \Biggr\} 
= -\frac{i H^2}{12 \pi^2} \, a^2 \delta^0_{\mu} \delta^0_{\nu} \; .
\end{equation}
Summing the three terms and multiplying by $-i 12 e^2 H^2 \delta \xi_{\rm 
fin}$ gives the total contribution from the conformal counterterm,
\begin{equation}
\frac{e^2 H^D \delta \xi_{\rm fin}}{(4 \pi)^{\frac{D}2}} \Gamma\Bigl(
\frac{D}2\Bigr) \Biggl\{ \Bigl[ \frac6{D \!-\!4} \!+\! 8\Bigr] g_{\mu\nu} - 
8 \, a^2 \delta^0_{\mu} \delta^0_{\nu} \Biggr\} . \label{ccount}
\end{equation}
In these and all subsequent formulae we neglect terms which vanish at 
$D\!=\!4$, or which fall off like $1/a$.

The procedure for $I^{1-6}_{\mu\nu}$ is the same. Consulting the appropriate
tables, acting derivatives and summing, we find,
\begin{eqnarray}
\lefteqn{I^1_{\mu\nu} = \frac{e^2 H^{2D-4}}{(4\pi)^D} \Biggl\{ \frac{(-D^2
\!+\! 6D \!-\! 6) \Gamma(D \!-\!1)}{(D \!-\! 2) (D\!-\!4)} \!-\! 9 \!+\! 
\pi^2\Biggr\} g_{\mu\nu} , } \\
\lefteqn{I^2_{\mu\nu} = \frac{e^2 H^{2D-4}}{(4\pi)^D} \Biggl\{ \Bigl[\frac{
(D^3 \!-\! 5 D^2 \!+\! 14 D \!-\! 14) \Gamma(D \!-\!1)}{(D\!-\!1) (D\!-\!2)
(D\!-\!3) (D\!-\!4)} \!+\! \pi^2 \Bigr] g_{\mu\nu} } \nonumber \\
& & \hspace{.7cm} + \Bigl[\frac{4 \Gamma(D \!-\!1)}{(D\!-\!1) (D\!-\!3)
(D\!-\!4)} \!-\! 8 \ln^2(a) \!+\! 32 \ln(a) \!-\! 58 \!+\! \frac83 \pi^2 
\Bigr] \, a^2 \delta_{\mu}^0 \delta_{\nu}^0 \Biggr\} , \qquad \\
\lefteqn{I^3_{\mu\nu} = \frac{e^2 H^{2D-4}}{(4\pi)^D} \Biggl\{ 
\frac{4 \Gamma(D \!-\!1)}{(D\!-\!2) (D\!-\!3) (D\!-\!4)} } \nonumber \\
& & \hspace{4.8cm} + 8 \ln^2(a) \!-\! 32 \ln(a) \!+\! 68 \!-\! \frac83 
\pi^2 \!\Biggr\} \, a^2 \delta_{\mu}^0 \delta_{\nu}^0 \; , \qquad \\
\lefteqn{I^4_{\mu\nu} = 0 \; , } \\
\lefteqn{I^5_{\mu\nu} = \frac{e^2 H^{2D-4}}{(4\pi)^D} \Biggl\{ 
8 \ln^2(a) \!-\! 32 \ln(a) \!+\! 48 \!-\! \frac43 \pi^2 \Biggr\} \, a^2 
\delta_{\mu}^0 \delta_{\nu}^0 , } \\
\lefteqn{I^6_{\mu\nu} = \frac{e^2 H^{2D-4}}{(4\pi)^D} \Biggl\{ 
-8 \ln^2(a) \!+\! 32 \ln(a) \!-\! 48 \!+\! \frac43 \pi^2 \Biggr\} \, a^2 
\delta_{\mu}^0 \delta_{\nu}^0 . }
\end{eqnarray}
The sum of all six integrals is,
\begin{eqnarray}
\lefteqn{\sum_{k=1}^6 I^k_{\mu\nu} = \frac{e^2 H^{2D-4}}{(4\pi)^D} \Biggl\{ 
\Bigl[\frac{(-D^3 \!+\! 9 D^2 \!-\! 20 D \!+\!16) \Gamma(D \!-\!1)}{(D\!-\!1) 
(D\!-\!3) (D\!-\!4)} \!-\! 9 \!+\! 2 \pi^2 \Bigr] g_{\mu\nu} } \nonumber \\
& & \hspace{3.8cm} + \Bigl[\frac{4 (2D \!-\!3) \Gamma(D \!-\!1)}{(D\!-\!1) 
(D\!-\!2) (D\!-3\!) (D\!-\!4)} \!+\! 10 \Bigr] \, a^2 \delta_{\mu}^0 
\delta_{\nu}^0 \Biggr\} . \qquad \label{sum6}
\end{eqnarray}
Note the cancellation of all infrared logarithms.

We can now substitute (\ref{ccount}) and (\ref{sum6}) in (\ref{total}) to
give our final result,
\begin{eqnarray}
\lefteqn{\Bigl\langle \Omega \Bigl\vert (D_{\mu} \varphi)^* D_{\nu} \varphi 
\Bigr\vert \Omega \Bigr\rangle \!=\! - \Bigl[1 \!-\! \delta Z_2\Bigr] 
\Bigl(\frac{D\!-\!1}{D}\Bigr) k H^2 g_{\mu\nu} \!+\! e^2 \gamma(0) \Bigl\{ A(0) 
\!+\! 2 k \ln(a) \Bigr\} g_{\mu\nu} } \nonumber \\
& & \hspace{-.5cm} + \frac{e^2 H^D \Gamma(\frac{D}2) \delta \xi_{\rm fin}}{(4 
\pi)^{\frac{D}2}} \Biggl\{ \Bigl[ \frac6{D \!-\!4} \!+\! 8\Bigr] g_{\mu\nu} 
- 8 \, a^2 \delta^0_{\mu} \delta^0_{\nu} \Biggr\} + \frac{e^2 H^{2D-4}}{(4
\pi)^D} \Biggl\{\Bigl[\frac{16 \Gamma(D \!-\!1)}{(D\!-\!1) (D\!-\!4)} 
\nonumber \\
& & - 17 \!+\! 2 \pi^2 \Bigr] g_{\mu\nu} + \Bigl[\frac{10
\Gamma(D \!-\!1)}{(D\!-\!1) (D\!-\!4)} \!+\! \frac83 \Bigr] \, a^2 
\delta_{\mu}^0 \delta_{\nu}^0 + O(D\!-\!4) \! \Biggr\} + O(e^4) \; . 
\qquad \label{final}
\end{eqnarray}
Note again that we have dropped terms which vanish for $D \!=\! 4$, and also
terms which fall off at late times, relative to the overall factor of $a^2$.
The only infrared logarithm in (\ref{final}) is precisely the one predicted
at order $e^2$ by the stochastic analysis \cite{PTsW1}.

\section{$\langle \Omega \vert \varphi^* \varphi \vert \Omega \rangle$}

It might seem that computing this VEV requires an analysis as extensive as
what we have just done. However, it is possible to get this result from the
previous one by making use of the exact Heisenberg equation of motion for
the scalar field operator,
\begin{equation}
\frac{(1 \!+\! \delta Z_2)}{\sqrt{-g}} D_{\mu} \Bigl(\sqrt{-g} g^{\mu\nu} 
D_{\nu} \varphi\Bigr) - \delta \xi R \varphi - \frac{\delta \lambda}2 \varphi
\varphi^* \varphi = 0 \; . \label{Heisen}
\end{equation}
Next act the scalar d`Alembertian on the gauge invariant product of 
$\varphi^*(x)$ and $\varphi(x)$, and substitute (\ref{Heisen}),
\begin{eqnarray}
\lefteqn{\square (\varphi^* \varphi) = \frac1{\sqrt{-g}} \Bigl[ D_{\mu} \Bigl(
\sqrt{-g} g^{\mu\nu} D_{\nu} \varphi\Bigr) \Bigr]^* \varphi } \nonumber \\
& & \hspace{3cm} + 2 g^{\mu\nu} (D_{\mu} \varphi)^* D_{\nu} \varphi + 
\varphi^* \frac1{\sqrt{-g}} D_{\mu} \Bigl(\sqrt{-g} g^{\mu\nu} D_{\nu} 
\varphi\Bigr) \; , \qquad \\
& & = 2 g^{\mu\nu} (D_{\mu} \varphi)^* D_{\nu} \varphi + \frac{2 \delta \xi R}{
1 \!+\! \delta Z_2} \varphi^* \varphi + \frac{\delta \lambda}{1 \!+\! \delta
Z_2} (\varphi^* \varphi)^2 \; . 
\end{eqnarray}
Because it is valid to use the equations of motion inside functional integrals
of gauge invariant operators \cite{TW8}, we conclude,
\begin{eqnarray}
\lefteqn{\square \Bigl\langle \Omega \Bigl\vert \varphi^* \varphi \Bigr\vert 
\Omega \Bigr\rangle = 2 g^{\mu\nu} \Bigl\langle \Omega \Bigl\vert (D_{\mu} 
\varphi)^* D_{\nu} \varphi \Bigr\vert \Omega \Bigr\rangle } \nonumber \\
& & \hspace{4cm} + \frac{2 \delta \xi R}{1 \!+\!  \delta Z_2} \Bigl\langle 
\Omega \Bigl\vert \varphi^* \varphi \Bigr\vert \Omega \Bigr\rangle + 
\frac{\delta \lambda}{1 \!+\! \delta Z_2} \Bigl\langle \Omega \Bigl\vert 
(\varphi^* \varphi)^2 \Bigr\vert \Omega \Bigr\rangle \; . \qquad
\end{eqnarray}
Finally, recall that $\delta \xi \sim e^2$, $\delta Z_2 \sim e^2$ and
$\delta \lambda \sim e^4$. Hence the result we require at one and two loop
orders is,
\begin{equation}
\square \Bigl\langle \Omega \Bigl\vert \varphi^* \varphi \Bigr\vert 
\Omega \Bigr\rangle = 2 g^{\mu\nu} \Bigl\langle \Omega \Bigl\vert (D_{\mu} 
\varphi)^* D_{\nu} \varphi \Bigr\vert \Omega \Bigr\rangle + 2 \delta \xi R
\Bigl\langle \Omega \Bigl\vert \varphi^* \varphi \Bigr\vert \Omega 
\Bigr\rangle + O(e^4) \; . \label{oureqn}
\end{equation}

To the order we are working, the last term in (\ref{oureqn}) is just $2 R
= 2 (D\!-\!1) D H^2$ times the product of the one loop conformal counterterm 
(\ref{delxi}) with the coincident scalar propagator,
\begin{equation}
2 \delta \xi R
\Bigl\langle \Omega \Bigl\vert \varphi^* \varphi \Bigr\vert \Omega \Bigr\rangle 
= \Bigl(-e^2  2 D \gamma(0) + 24 e^2 H^2 \delta \xi_{\rm fin}\Bigr)
\Bigl(A(0) + 2 k \ln(a)\Bigr) + O(e^4) \; .
\end{equation}
The first term on the right hand side of (\ref{oureqn}) is just twice the
trace of our result (\ref{final}) from the previous section,
\begin{eqnarray}
\lefteqn{2 g^{\mu\nu} \Bigl\langle \Omega \Bigl\vert (D_{\mu} \varphi)^* 
D_{\nu} \varphi \Bigr\vert \Omega \Bigr\rangle = - 2D \Bigl[1 \!-\! 
\delta Z_2\Bigr] \Bigl(\frac{D\!-\!1}{D}\Bigr) k H^2 } \nonumber \\
& & + e^2 2 D \gamma(0) \Bigl\{ A(0) \!+\! 2 k \ln(a) \Bigr\} + \frac{e^2 H^D 
\delta \xi_{\rm fin}}{(4\pi)^{\frac{D}2}} \Gamma\Bigl(\frac{D}2\Bigr) 
\Bigl\{\frac{12 D}{D\!-\!4} \!+\! 80\Bigr\} \nonumber \\
& & \hspace{1cm} + \frac{e^2 H^{2D-4}}{(4\pi)^D} \Biggl\{\frac{108
\Gamma(D \!-\! 1)}{(D\!-\!1)(D\!-\!4)} \!-\! 120 \!+\! 16 \pi^2 \!+\!
O(D\!-\!4)\Biggr\} \!+\! O(e^4) \; . \qquad
\end{eqnarray}
Substituting these two relations in (\ref{oureqn}) results in complete 
cancellation of the divergent infrared logarithms to the order we are working,
\begin{eqnarray}
\lefteqn{\square \Bigl\langle \Omega \Bigl\vert \varphi^* \varphi \Bigr\vert 
\Omega \Bigr\rangle = - 2D \Bigl[1 \!-\! \delta Z_2\Bigr] 
\Bigl(\frac{D\!-\!1}{D}\Bigr) k H^2 } \nonumber \\
& & + 24 e^2 H^2 \delta \xi_{\rm fin} \Bigl\{ A(0) \!+\! 2 k \ln(a) \Bigr\} + 
\frac{e^2 H^D \delta \xi_{\rm fin}}{(4\pi)^{\frac{D}2}} \Gamma\Bigl(\frac{D}2 
\Bigr) \Bigl\{\frac{12 D}{D\!-\!4} \!+\! 80\Bigr\} \nonumber \\
& & \hspace{1cm} + \frac{e^2 H^{2D-4}}{(4\pi)^D} \Biggl\{\frac{108
\Gamma(D\!-\!1)}{(D\!-\!1)(D\!-\!4)} \!-\! 120 \!+\! 16 \pi^2 \!+\!
O(D\!-\!4) \Biggr\} \!+\! O(e^4) \; . \qquad
\end{eqnarray}
In fact we can eliminate any infrared logarithms, at order $e^2$, by choosing 
the finite part of the conformal counterterm to vanish,
\begin{eqnarray}
\lefteqn{\square \Bigl\langle \Omega \Bigl\vert \varphi^* \varphi \Bigr\vert 
\Omega \Bigr\rangle \Biggl\vert_{\delta \xi_{\rm fin} = 0} = - 2D \Bigl[1 \!-\! 
\delta Z_2\Bigr] \Bigl(\frac{D\!-\!1}{D}\Bigr) k H^2 } \nonumber \\
& & \hspace{1cm} + \frac{e^2 H^{2D-4}}{(4\pi)^D} \Biggl\{\frac{108
\Gamma(D\!-\!1)}{(D\!-\!1)(D\!-\!4)} \!-\! 120 \!+\! 16 \pi^2 \!+\! O(D\!-\!4)
\Biggr\} \!+\! O(e^4) \; . \qquad \label{2ndpre}
\end{eqnarray}

Note that on a function of $\ln(a)$, the scalar d`Alembertian gives,
\begin{equation}
\square f\Bigl(\ln(a)\Bigr) = -H^2 \Biggl\{ f''\Bigl(\ln(a)\Bigr) +
(D\!-\!1) f'\Bigl(\ln(a)\Bigr) \Biggr\} \; .
\end{equation}
If the d`Alembertian of such a function is a constant $K$ then we can
reconstruct the function up to an integration constant,
\begin{equation}
\square f\Bigl(\ln(a)\Bigr) = K \qquad \Longrightarrow \qquad 
f\Bigl(\ln(a)\Bigr) = -\frac{K \ln(a)}{(D\!-\!1) H^2} + {\rm constant} \; .
\end{equation}
We see from this and (\ref{2ndpre}) that choosing $\delta \xi_{\rm fin} = 0$ 
results in the VEV of $\varphi^*(x) \varphi(x)$ possessing no $\ln^2(a)$ 
contribution at order $e^2$. It is interesting to note that $\delta \xi_{\rm 
fin} = 0$ is also the unique choice which results in there being no 
significant late time corrections to the scalar mode functions at one loop
order \cite{KW2}.

At order $e^0$ the VEV of $\varphi^*(x) \varphi(x)$ is just the coincident
propagator, and it contains a single infrared logarithm. Hence the leading
logarithm correction for this VEV at order $e^2$ would contain two infrared
logarithms. We have just seen that choosing $\delta \xi_{\rm fin} = 0$ 
causes this leading logarithm correction to vanish. That is another key
prediction of the stochastic formalism \cite{PTsW1}.

\section{Discussion}

We have used dimensional regularization to compute the one and two loop 
VEV's of two gauge invariant operators in SQED. Our results (\ref{final})
and (\ref{2ndpre}) confirm two predictions of the stochastic analysis 
\cite{PTsW1}:
\begin{itemize}
\item{That the leading log result for the coincident kinetic term is,
\begin{eqnarray}
\lefteqn{\Bigl\langle \Omega \Bigl\vert (D_{\mu} \varphi)^* D_{\nu} \varphi 
\Bigr\vert \Omega \Bigr\rangle_{\rm LL} } \nonumber \\
& & \hspace{1cm} = \frac{H^4 g_{\mu\nu}}{16 \pi^2} \Biggl\{-\frac32 + 
\frac3{D\!-\!4} \times \frac{e^2}{4 \pi^2} \ln(a) + O\Bigl(e^4 \ln^2(a)\Bigr) 
\Biggr\} . \qquad \label{pas}
\end{eqnarray}}
\item{That setting the conformal counterterm to,
\begin{equation}
\delta \xi = \frac{e^2 H^{D-4}}{(4\pi)^{\frac{D}2}} \Biggl\{ -\frac1{D\!-\!4}
+ \frac{\gamma}2 + O(D\!-\!4) \Biggr\} , \label{spconf}
\end{equation}
(which corresponds to $\delta \xi_{\rm fin} = 0$) results in the coincident 
scalar norm having no leading logs at order $e^2$,
\begin{equation}
\Bigl\langle \Omega \Bigl\vert \varphi^*(x) \varphi(x) \Bigr\vert \Omega 
\Bigr\rangle_{\rm LL} = \frac{H^2 \ln(a)}{4 \pi^2} \Biggl\{1 + 0 \times
\frac{e^2}{4 \pi^2} \ln(a) + O\Bigl(e^4 \ln^2(a)\Bigr) \Biggr\} . \label{act}
\end{equation}}
\end{itemize}

Two additional points deserve comment concerning the leading logarithm
approximation. First, we saw in Yukawa theory \cite{MW3} that leading 
logarithm corrections to the VEV's of passive fields can harbor ultraviolet 
divergences. From (\ref{pas}) we see that the same can be true for leading 
logarithm corrections to the VEV's of differentiated active fields. In both 
cases the reason is that the ultraviolet cannot be ignored in any field which
fails to contribute an infrared logarithm. By contrast, the VEV of 
undifferentiated active fields such as (\ref{act}) must be finite at leading 
logarithm order. Note that this would be true no matter what choice had been 
made for $\delta \xi_{\rm fin}$.

Our second observation is that choosing (\ref{spconf}) prevents significant 
late time corrections to the scalar mode functions \cite{KW2} at one loop.
This seems to be the unique renormalization prescription which suppresses
quantum secular effects as much as possible at order $e^2$. However, there
is no way to prevent significant corrections to the photon mode functions at 
one loop order \cite{PTW1,PTW2,PW1}.

Although the principal application of this exercise has been to provide
``data'' for checking testing a leading-log resummation of SQED \cite{PTsW1},
the calculation is not without interest in its own right. Because both
the operators whose VEV's we computed are gauge invariant, it would not have
mattered which gauge we used. However, we found that working in Lorentz gauge
(\ref{gauge}) greatly simplified the one loop self-mass-squared. It also
permitted the two loop scalar kinetic operator (\ref{pas}) to be expressed 
in terms of a single-vertex integral, rather than the 2-vertex integration 
that seems to be required for the leftmost diagram of Fig.~3.

Another significant technical advance (for which see the Appendix) is that 
we have worked out procedures for integrating functions of $y(x;x')$ by
extracting covariant d`Alembertians. This has advantages over the technique 
of breaking $y(x;x') = H^2 a a' \Delta x^2$ up into factors and then
extracting powers of the flat space d`Alembertian $\partial^2 = \eta^{\mu\nu} 
\partial_{\mu} \partial_{\nu}$. The older technique was employed in all 
previous computations of this sort \cite{OW1,OW2,BOW,PTW2,PW3,MW1,KW,DW}.
Although it produces correct results, one must work through many tedious 
cancellations between spurious infrared logarithms from the $D$-dependent 
powers of $a$ which reside on ultraviolet divergences, and equally spurious 
infrared logarithms from the nonlocal, finite terms. The new technique
organizes the calculation so that these spurious infrared logarithms never
appear in the first place.

\section{Appendix: Tables~\ref{dint},~\ref{f4int} and~\ref{f3int}}

The purpose of this appendix is to describe how to evaluate certain
integrals of the form,
\begin{equation}
\int d^Dx' a^{\prime D} \Bigl\{f(y_{\scriptscriptstyle ++}) \!-\! f(y_{
\scriptscriptstyle +-})\Bigr\} \quad {\rm and} \quad
\int d^Dx' a^{\prime D-1} \Bigl\{f(y_{\scriptscriptstyle ++}) \!-\! f(y_{
\scriptscriptstyle +-})\Bigr\} \; . \qquad
\end{equation}
Section 3 has already discussed our method for segregating ultraviolet 
divergences on to delta functions. Briefly, the procedure is to extract 
covariant d`Alembertians using the identity,
\begin{equation}
\Bigl(\frac{4}{y}\Bigr)^{\alpha} = \frac1{(\alpha \!-\! 1)(\alpha \!-\! 
\frac{D}2)} \frac{\square}{H^2} \Bigl(\frac{4}{y}\Bigr)^{\alpha -1}
+ \Bigl(\frac{\alpha \!-\! D}{\alpha \!-\! \frac{D}2}\Bigr) \Bigl(\frac{4}{y}
\Bigr)^{\alpha -1} \qquad \forall \, \alpha \neq \frac{D}2 \; .
\end{equation}
One continues applying this until the singularities $1/y^{\alpha-1}$ are
integrable for $D \!=\! 4$. At this stage the denominator $(\alpha \!-\! 
\frac{D}2)$ vanishes for $D\!=\! 4$. To segregate this divergence on a local 
term one adds zero in the form,
\begin{eqnarray}
0 & = & \frac{(4\pi)^{\frac{D}2} H^{-D}}{\Gamma(\frac{D}2 \!-\! 1)}
\frac{i \delta^D(x \!-\! x')}{\sqrt{-g}} - \frac{\square}{H^2} 
\Bigl(\frac{4}{y_{\scriptscriptstyle ++}} \Bigr)^{\frac{D}2-1} + \frac{D}2 
\Bigl(\frac{D}2 \!-\! 1\Bigr) \Bigl(\frac{4}{y_{\scriptscriptstyle ++}}
\Bigr)^{\frac{D}2 -1} \; , \qquad \\
0 & = & 0 - \frac{\square}{H^2} \Bigl(\frac{4}{y_{\scriptscriptstyle +-}} 
\Bigr)^{\frac{D}2-1} + \frac{D}2 \Bigl(\frac{D}2 \!-\! 1\Bigr) 
\Bigl(\frac{4}{y_{\scriptscriptstyle +-}} \Bigr)^{\frac{D}2 -1} \; . \qquad
\end{eqnarray}
One then takes $D \!\rightarrow\! 4$ in the nonlocal terms.

The results for the two potentially divergent integrands we require are,
\begin{eqnarray}
\lefteqn{\Bigl(\frac{4}{y_{\scriptscriptstyle ++}}\Bigr)^{D-2} = \frac2{(D\!-\!
3)(D\!-\!4)} \frac{(4\pi)^{\frac{D}2} H^{-D}}{\Gamma(\frac{D}2 \!-\!1)}
\frac{i \delta^D(x \!-\!x')}{\sqrt{-g}} - \frac{\square}{H^2} \Biggl\{\Bigl(
\frac{4}{y_{\scriptscriptstyle ++}}\Bigr) \ln\Bigl(\frac{y_{\scriptscriptstyle 
++}}4\Bigr)\Biggr\} } \nonumber \\
& & \hspace{5cm} + 2 \Bigl(\frac{4}{y_{\scriptscriptstyle ++}}\Bigr)
\ln\Bigl(\frac{y_{\scriptscriptstyle ++}}4\Bigr) -
\frac{4}{y_{\scriptscriptstyle ++}} + O(D\!-\!4) \; , \label{yD-2} \qquad \\
\lefteqn{\Bigl(\frac{4}{y_{\scriptscriptstyle ++}}\Bigr)^{D-1} = \frac2{(D\!-\!
2)^2} \frac{\square}{H^2} \Bigl(\frac4{y_{\scriptscriptstyle ++}}\Bigr)^{D-2}
-\frac2{D\!-\!2} \Bigl(\frac{4}{y_{\scriptscriptstyle ++}}\Bigr)^{D-2} \; . }
\label{1stD-1}
\end{eqnarray}
The $+-$ results follow from these by dropping the delta functions and
replacing $y_{\scriptscriptstyle ++}$ everywhere with
$y_{\scriptscriptstyle +-}$.

Of course it is quite simple to evaluate integrals of delta functions!
The finite, nonlocal terms require an additional partial integration
to remove factors of $1/y$. The two identities we need are,
\begin{eqnarray}
\frac4{y} & = & \frac{\square}{H^2} \Biggl\{ \ln\Bigl(\frac{y}4\Bigr)\Biggr\}
+ 3 \; , \label{1/x} \qquad \\
\frac4{y} \ln\Bigl(\frac{y}4\Bigr) & = & \frac{\square}{H^2} \Biggl\{ \frac12
\ln^2\Bigl(\frac{y}4\Bigr) \!-\! \ln\Bigl(\frac{y}4\Bigr) \Biggr\}
+ 3\ln\Bigl(\frac{y}4\Bigr) - 2 \; . \label{ln/x} \qquad
\end{eqnarray}
These d`Alembertians are extracted from the integration over $x^{\prime \mu}$
and then acted after performing the integrals. Because the integrals can only
depend upon $a$, the following identities are useful for acting the
d`Alembertians,
\begin{eqnarray}
\frac{\square}{H^2} \Bigl(1\Bigr) = 0 & , & \frac{\square}{H^2} 
\Bigl(\frac1{a}\Bigr) = \frac2{a} \; , \label{dln0} \\
\frac{\square}{H^2} \Bigl( \ln(a) \Bigr) = -3 & , & \frac{\square}{H^2} 
\Bigl(\frac{\ln(a)}{a}\Bigr) = \frac{2 \ln(a)}{a} \!-\! \frac1{a} \; , 
\label{dln1} \\
\frac{\square}{H^2} \Bigl( \ln^2(a) \Bigr) = -6 \ln(a) \!-\! 2 & , & 
\frac{\square}{H^2} \Bigl(\frac{\ln^2(a)}{a}\Bigr) = \frac{2 \ln^2(a)}{a} \!-\!
\frac{2 \ln(a)}{a} \!-\! \frac2{a} \; . \qquad \label{dln2}
\end{eqnarray}

We will shortly describe how to perform the finite integrals which make up
Tables~\ref{f4int} and \ref{f3int} but it is best to first complete the
discussion of the potentially divergent integrals of Table~\ref{dint}.
Combining (\ref{yD-2}) with (\ref{1/x}) and (\ref{ln/x}), we first express 
the integral as a divergent part plus a sum of differentiated finite
integrals. These finite integrals are performed using Tables~\ref{f4int}
and \ref{f3int}, then the derivatives are acted using relations 
(\ref{dln0}-\ref{dln2}). The last step involves expanding some terms on the 
divergent part to obtain a simpler answer for tabulation. The steps for the 
bottom left entry in Table~\ref{dint} are,
\begin{eqnarray}
\lefteqn{\int d^Dx' a^{\prime D} \Biggl\{ \Bigl(\frac{4}{y_{\scriptscriptstyle 
++}}\Bigr)^{D-2} \!-\! \Bigl(\frac{4}{y_{\scriptscriptstyle +-}}\Bigr)^{D-2}
\Biggr\} = \frac{2}{(D\!-\!3) (D\!-\!4)} \frac{i (4\pi)^{\frac{D}2}}{ H^D 
\Gamma(\frac{D}2 \!-\! 1)} } \nonumber \\
& & \hspace{-.5cm} - \frac{\square^2}{H^4} \int d^4x' a^{\prime 4} \Biggl\{
\frac12 \ln^2\Bigl(\frac{y_{\scriptscriptstyle ++}}4\Bigr) \!-\! \ln\Bigl(
\frac{y_{\scriptscriptstyle ++}}4\Bigr) - (+-)\Biggr\} + \frac{\square}{H^2} 
\int d^4x' a^{\prime 4} \Biggl\{\ln^2\Bigl(\frac{y_{\scriptscriptstyle ++}}4
\Bigr) \nonumber \\
& & - 6 \ln\Bigl(\frac{y_{\scriptscriptstyle ++}}4\Bigr) \!-\! 
(+-)\Biggr\} \!+\! \int d^4x' a^{\prime 4} \Biggl\{ 6 \ln\Bigl(\frac{y_{
\scriptscriptstyle ++}}4 \Bigr) \!-\! (+-)\Biggr\} \!+\! O(D\!-\!4) \; , 
\qquad \\
& & \hspace{-.5cm} = \frac{i (4\pi)^{\frac{D}2}}{H^D \Gamma(\frac{D}2 \!-\!1)}
\Biggl\{ \frac{2}{(D\!-\!3)(D\!-\!4)} -\frac{\square^2}{H^4} 
\left[{\frac1{12} \ln^2(a) - \frac49 \ln(a) + \frac78 - \frac{\pi^2}{18} \atop
-\frac16 \ln(a) + \frac{11}{36}}\right] \nonumber \\
& & + \frac{\square}{H^2} \left[{\frac16 \ln^2(a) - \frac89 \ln(a) + \frac74
-\frac{\pi^2}9 \atop -\ln(a) + \frac{11}6}\right] + \Bigl[\ln(a) \!-\!
\frac{11}6\Bigr] + O(D\!-\!4) \Biggr\} , \qquad \\
& & \hspace{-.5cm} = \frac{i (4\pi)^{\frac{D}2}}{H^D \Gamma(\frac{D}2 \!-\!1)}
\Biggl\{ \frac{2}{(D\!-\!3)(D\!-\!4)} + 2 + O(D\!-\!4) \Biggr\} , \\
& & \hspace{-.5cm} = \frac{i (4\pi)^{\frac{D}2}}{H^D \Gamma(\frac{D}2 \!-\!1)}
\Biggl\{ \frac{2}{D\!-\!4} + O(D\!-\!4) \Biggr\} . \label{D-2/1}
\end{eqnarray}
The analogous $a^{\prime D-1}$ term gives,
\begin{eqnarray}
\lefteqn{\int d^Dx' a^{\prime D-1} \Biggl\{\Bigl(\frac{4}{y_{\scriptscriptstyle 
++}}\Bigr)^{D-2} \!-\! \Bigl(\frac{4}{y_{\scriptscriptstyle +-}}\Bigr)^{D-2}
\Biggr\} = \frac{2}{(D\!-\!3) (D\!-\!4)} \frac{i (4\pi)^{\frac{D}2} a^{-1}}{ 
H^D \Gamma(\frac{D}2 \!-\! 1)} } \nonumber \\
& & \hspace{-.5cm} - \frac{\square^2}{H^4} \int d^4x' a^{\prime 3} \Biggl\{
\frac12 \ln^2\Bigl(\frac{y_{\scriptscriptstyle ++}}4\Bigr) \!-\! \ln\Bigl(
\frac{y_{\scriptscriptstyle ++}}4\Bigr) - (+-)\Biggr\} + \frac{\square}{H^2} 
\int d^4x' a^{\prime 3} \Biggl\{\ln^2\Bigl(\frac{y_{\scriptscriptstyle ++}}4
\Bigr) \nonumber \\
& & - 6 \ln\Bigl(\frac{y_{\scriptscriptstyle ++}}4\Bigr) \!-\! 
(+-)\Biggr\} \!+\! \int d^4x' a^{\prime 3} \Biggl\{ 6 \ln\Bigl(\frac{y_{
\scriptscriptstyle ++}}4 \Bigr) \!-\! (+-)\Biggr\} \!+\! O(D\!-\!4) \; , 
\qquad \\
& & \hspace{-.5cm} = \frac{i (4\pi)^{\frac{D}2}}{H^D \Gamma(\frac{D}2 \!-\!1)}
\Biggl\{ \frac{2 a^{-1}}{(D\!-\!3)(D\!-\!4)} -\frac{\square^2}{H^4} 
\left[{\frac16 \ln(a) - \frac{11}{18} - \frac{\ln^2(a)}{4 a} + \frac{\ln(a)}{a}
\atop - \frac{31}{24 a} + \frac{\pi^2}{6 a} 
-\frac16 + \frac{\ln(a)}{2 a} - \frac1{4a}}\right] \nonumber \\
& & + \frac{\square}{H^2} \left[{\frac13 \ln(a) - \frac{11}9  - 
\frac{\ln^2(a)}{2 a} + \frac{2 \ln(a)}{a} - \frac{31}{12 a} + \frac{\pi^2}{3 a}
\atop -1 + \frac{3 \ln(a)}{a} - \frac{3}{2 a}}\right] \nonumber \\
& & \hspace{6cm} + \Bigl[1 \!-\! \frac{3 \ln(a)}{a} + \frac{3}{2 a} \Bigr] 
+ O(D\!-\!4) \Biggr\} , \qquad \\
& & \hspace{-.5cm} = \frac{i (4\pi)^{\frac{D}2} }{H^D \Gamma(\frac{D}2 
\!-\!1)} \Biggl\{ \frac{2 a^{-1}}{(D\!-\!3)(D\!-\!4)} + O(D\!-\!4)\Biggr\} , \\
& & \hspace{-.5cm} = \frac{i (4\pi)^{\frac{D}2} }{H^D \Gamma(\frac{D}2\!-\!1)} 
\Biggl\{ \frac{2 a^{-1}}{D\!-\!4} - \frac{2}{a} + O(D\!-\!4) \Biggr\}.
\label{D-2/a}
\end{eqnarray}
The top two entries in Table~{\ref{dint}} follow from using relation 
(\ref{1stD-1}) on (\ref{D-2/1}) and (\ref{D-2/a}), respectively.

We turn now to the finite integrals in Tables~\ref{f4int} and \ref{f3int}.
First note that relations (\ref{1/x}) and (\ref{ln/x}) allow us to express
the results for $f_1(x) \!=\! 1/x$ and $f_2(x) \!=\! \ln(x)/x$ in terms of 
$f_7(x) \!=\! \ln(x)$ and $f_8(x) \!=\! \ln^2(x)$,
\begin{eqnarray}
f_1\Bigl(\frac{y_{\scriptscriptstyle ++}}4\Bigr) -
f_1\Bigl(\frac{y_{\scriptscriptstyle +-}}4\Bigr) & = &
\frac{\square}{H^2} \Biggl\{f_7\Bigl(\frac{y_{\scriptscriptstyle ++}}4\Bigr) -
f_7\Bigl(\frac{y_{\scriptscriptstyle +-}}4\Bigr) \Biggr\} , \qquad \\
f_2\Bigl(\frac{y_{\scriptscriptstyle ++}}4\Bigr) -
f_2\Bigl(\frac{y_{\scriptscriptstyle +-}}4\Bigr) & = &
\frac{\square}{H^2} \Biggl\{\frac12 f_8\Bigl(\frac{y_{\scriptscriptstyle ++}}4
\Bigr) - f_7\Bigl(\frac{y_{\scriptscriptstyle ++}}4\Bigr) - (+-) \Biggr\} 
\nonumber \\
& & \hspace{3cm} + 3 \Biggr\{f_7\Bigl(\frac{y_{\scriptscriptstyle ++}}4\Bigr) -
f_7\Bigl(\frac{y_{\scriptscriptstyle +-}}4\Bigr) \Biggr\} . \qquad
\end{eqnarray}
So one uses these relations, and the d`Alembertian identities 
(\ref{dln0}-\ref{dln2}), to derive the first two entries on Tables~\ref{f4int} 
and \ref{f3int} from the bottom two entries.

The procedure for integrating $f_7(x)$ and $f_8(x)$ is straightforward:
\begin{itemize}
\item{Combine the $++$ and $+-$ terms to extract a factor of $i$ and to
make causality manifest,
\begin{eqnarray}
\ln\Bigl(\frac{y_{\scriptscriptstyle ++}}4\Bigr) - 
\ln\Bigl(\frac{y_{\scriptscriptstyle +-}}4\Bigr) & = & 2 \pi i 
\theta\Bigl(\Delta \eta \!-\! \vert \vec{x} \!-\! \vec{x}'\vert\Bigr)
\; , \qquad \\
\ln^2\Bigl(\frac{y_{\scriptscriptstyle ++}}4\Bigr) - 
\ln^2\Bigl(\frac{y_{\scriptscriptstyle +-}}4\Bigr) & = & 4 \pi i 
\theta\Bigl(\Delta \eta \!-\! \vert \vec{x} \!-\! \vec{x}'\vert\Bigr)
\ln\Bigl(-y(x;x')\Bigr) \; . \qquad
\end{eqnarray}
Here we define $\Delta \eta \equiv \eta - \eta'$.}
\item{Make the change of variables $\vec{r} = \vec{x}' - \vec{x}$ and perform
the angular integrations,
\newpage
\begin{eqnarray}
\lefteqn{\int d^3x' \theta\Bigl(\Delta \eta \!-\! \vert\vec{x} \!-\! \vec{x}'
\vert\Bigr) F\Bigl(-y(x;x')\Bigr) } \nonumber \\
& & \hspace{2cm} = 4\pi \theta(\Delta \eta) \int_0^{\Delta \eta} \!\! dr r^2 
F\Bigl(a a' H^2 (\Delta \eta^2 \!-\! r^2)\Bigr) \; . \qquad
\end{eqnarray}}
\item{Make the change of variables $r = \Delta \eta \cdot z$ and perform the
integration over $z$.}
\item{Make the change of variables $a' = -1/H\eta'$ and perform the
integration over $a'$.}
\item{Discard terms which fall off at late times with respect to $1$ for 
Table~\ref{f4int}, or to $1/a$ for Table~\ref{f3int}.}
\end{itemize}
For the case of $f_8(x)$ integrated against $a^{\prime 4}$, the steps are,
\begin{eqnarray}
\lefteqn{\int d^4x' a^{\prime 4} \Biggl\{ \ln^2\Bigl(\frac{y_{
\scriptscriptstyle ++}}4\Bigr) - \ln^2\Bigl(\frac{y_{\scriptscriptstyle +-}}4
\Bigr) \Biggr\} } \nonumber \\
& & = 4\pi i \times 4\pi \!\! \int_{\eta_i}^{\eta} \!\! d\eta' a^{\prime 4} 
\Delta \eta^3 \!\! \int_0^1 \!\! dz z^2 \Biggl\{\ln(a a') \!+\! 2 \ln(H\Delta 
\eta) \!+\! \ln\Bigl(\frac{1 \!-\! z^2}4\Bigr) \Biggr\} , \qquad \\
& & = \frac{16 \pi^2 i}{H^4} \!\! \int_1^a \!\! da' a^{\prime 2} \Bigl(
\frac1{a'} \!-\! \frac1{a}\Bigr)^3 \Biggl\{-\frac13 \ln\Bigl(\frac{a'}{a}\Bigr)
\!+\! \frac23 \ln\Bigl(1 \!-\! \frac{a'}{a}\Bigr) \!-\! \frac89\Biggr\} , 
\qquad \\
& & = \frac{16 \pi^2 i}{H^4} \Biggl\{\frac16 \ln^2(a) \!-\! \frac89 \ln(a)
\!+\! \frac{7}4 \!-\! \frac{\pi^2}{9} \!+\! \Bigl[\frac1{a} \!-\! \frac1{2a^2}
\!+\! \frac1{9 a^3}\Bigr] \ln(a) \!+\! \frac1{9a} \nonumber \\
& & \hspace{1.2cm} + \frac{13}{12 a^2} \!-\! \frac{7}{27 a^3}
\!+\! 2 \sum_{n=1}^{\infty} \Bigl[ \frac{a^{-n}}{3 n^2} \!-\! \frac{a^{-n-1}}{
n (n\!+\!1)} \!+\! \frac{a^{-n-2}}{n (n\!+\!2)} \!-\! \frac{a^{-n-3}}{3 n
(n\!+\!3)}\Bigr] \Biggr\} , \qquad \\
& & = \frac{16 \pi^2 i}{H^4} \Biggl\{\frac16 \ln^2(a) \!-\! \frac89 \ln(a)
\!+\! \frac{7}4 \!-\! \frac{\pi^2}{9} \!+\! O\Bigl(\frac{\ln(a)}{a}\Bigr)
\Biggr\} .
\end{eqnarray}
Note that it is only at the last step for which it matters whether the measure
factor is $a^{\prime 4}$ or $a^{\prime 3}$.

The procedure is almost the same for $f_3(x) \!=\! \ln(x)/(1\!-\!x)^2$ and
$f_4(x) \!=\! \ln(x)/(1\!-\!x)$, with two exceptions. First, one extends the
range of $z$ and makes the change of variable,
\begin{equation}
x = \frac12 \Bigl[\frac1{a'} \!+\! \frac1{a} + \Bigl(\frac1{a'} \!-\!
\frac1{a}\Bigr) z\Bigr] \qquad \Longleftrightarrow \qquad z = \frac{2x \!-\! 
\frac1{a'} \!-\! \frac1{a}}{\frac1{a'} \!-\! \frac1{a}} \; .
\end{equation}
This carries the factor of $\Delta \eta^3$ times the $z$ integral to,
\begin{eqnarray}
\lefteqn{\Bigl(\frac1{a'} \!-\! \frac1{a}\Bigr)^3 \int_0^1 \!\!dz z^2 F\Biggl(1 
\!+\! a a' \Bigl(\frac1{a'} \!-\! \frac1{a}\Bigr)^2 \Bigl(\frac{1 \!-\! z^2}4
\Bigr)\Biggr) } \nonumber \\
& & \hspace{4cm} = \int_{\frac1{a}}^{\frac1{a'}} \!\! dx \Bigl(2x \!-\! 
\frac1{a'} \!-\! \frac1{a}\Bigr)^2 F\Biggl(a a' x \Bigl(\frac1{a'} \!+\! 
\frac1{a} \!-\! x\Bigr)\Biggr) \; . \qquad
\end{eqnarray}
The second exception is that one changes variables in the temporal integration
from $a'$ to $\alpha \!=\! a'/a$. For $f_3(x)$ integrated against $a^{\prime 
3}$, the steps are,
\begin{eqnarray}
\lefteqn{\int d^4x' a^{\prime 3} \Biggl\{ \frac{\ln(\frac{y_{++}}4)}{(1 \!-\!
\frac{y_{++}}4)^2} - \frac{\ln(\frac{y_{+-}}4)}{(1 \!-\! \frac{y_{+-}}4)^2}
\Biggr\} } \nonumber \\
& & \hspace{1.5cm} = \frac{2\pi i \times 4\pi}{H^4} \int_1^{a} \! da' a' 
\int_0^1 \!\! dz z^2 \frac1{[1 \!+\! a a' (\frac1{a'} \!-\! \frac1{a})^2 
(\frac{1 \!-\! z^2}4)]^2} \; , \qquad \\
& & \hspace{1.5cm} = \frac{8 \pi^2 i}{H^4} \int_1^{a} \! da' a' \times 
\frac1{(a a')^2} \int_{\frac1{a}}^{\frac1{a'}} \!\! dx \frac{[2x \!-\! 
(\frac1{a'} \!+\! \frac1{a})]^2}{[x (\frac1{a'} \!+\! \frac1{a} \!-\! x)]^2} 
\; , \qquad \\
& & \hspace{1.5cm} = \frac{8 \pi^2 i}{H^4} \int_1^{a} \! da' a' \times 
\frac2{(a a')^2} \Biggl\{-a' \!+\! a \!+\! \frac{2 \ln(\frac{a'}{a})}{
\frac1{a'} \!+\! \frac1{a}} \Biggr\} \; , \qquad \\
& & \hspace{1.5cm} = \frac{16 \pi^2 i}{H^4 a} \int_{\frac1{a}}^1 \!\!\ d\alpha 
\Biggl\{ -1 \!+\! \frac1{\alpha} \!+\! \frac{2 \ln(\alpha)}{1 \!+\! \alpha} 
\Biggr\} \; , \qquad \label{insert} \\
& & \hspace{1.5cm} = \frac{16 \pi^2 i}{H^4 a} \Biggl\{-\alpha \!+\! \ln(\alpha) 
\!+\! 2 \sum_{n=0}^{\infty} \frac{(-1)^n \alpha^{n+1}}{n\!+\!1} \Bigl[ 
\ln(\alpha) \!-\! \frac1{n\!+\!1}\Bigr] 
\Biggr\} \Biggr\vert_{\frac1{a}}^1 \; , \qquad \\
& & \hspace{1.5cm} = \frac{16 \pi^2 i}{H^4 a} \Biggl\{\ln(a) \!-\! 1 \!-\! 
\frac{\pi^2}{6} \!+\! O\Bigl(\frac{\ln(a)}{a}\Bigr)\Biggr\} \; . \qquad 
\end{eqnarray}
Note that the $a^{\prime 4}$ result (on Table~\ref{f4int}) would follow from
simply multiplying the integrand of (\ref{insert}) by $a \alpha$.

The most difficult reductions are those for $f_5(x) \!=\! \ln^2(x) \!-\! 2
\ln(1 \!-\! x) \ln(x)$ and $f_6(x) \!=\! x \ln^2(x) \!-\! 2 x \ln(1 \!-\! x)
\ln(x) \!-\! 2\ln(x)$. These combinations of factors were chosen because they
arise in performing the various integrations and multiplications, and because
cancellations between the different factors prevent significant contributions 
from the limit at $\alpha \!=\! 1$. We carry out the same reduction as for 
$f_3(x)$ and $f_4(x)$ on each factor separately, up to the $\alpha$
integration, and then combine them to take advantage of the cancellations.
The two factors in $f_5(x)$ times $a^{\prime 4}$ produce,
\begin{eqnarray}
\ln^2(x) & \!\!\! \Longrightarrow \!\!\! & \frac{16 \pi^2 i}{H^4} \!\!
\int_{\frac1{a}}^1 \! \frac{d\alpha}{\alpha} (1 \!-\! \alpha)^3 
\Bigl\{-\frac13 \ln(\alpha) +\frac23 \ln(1 \!-\! \alpha) \!-\! \frac89\Bigr\} ,
\qquad \\
-2\ln(1\!-\!x) \ln(x) & \!\!\! \Longrightarrow \!\!\! & \frac{16 \pi^2 i}{H^4} 
\!\! \int_{\frac1{a}}^1 \! \frac{d\alpha}{\alpha} \Bigl\{\frac13 (1 \!+\! 
\alpha)^3 \ln(\alpha) \!+\! \frac89 \!-\! \frac89 \alpha^3 \Bigr\} . \qquad
\end{eqnarray}
These two factors sum to give the full result for $f_5(x)$ times 
$a^{\prime 4}$,
\begin{eqnarray}
\lefteqn{\frac{16 \pi^2 i}{H^4} \int_{\frac1{a}}^1 d\alpha \Biggl\{
\Bigl[2 \!+\! \frac{2 \alpha^2}3 \Bigr] \ln(\alpha) 
+\frac2{3 \alpha} (1 \!-\! \alpha)^3 \ln(1\!-\! \alpha)
\!+\! \frac83 \!-\! \frac{8\alpha}3 \Biggr\} } \nonumber \\
& & \hspace{4cm} = \frac{16 \pi^2 i}{H^4} \Biggl\{ \frac16 - \frac{\pi^2}9 + 
O\Bigl(\frac{\ln(a)}{a}\Bigr) \Biggr\} . \qquad
\end{eqnarray}
Of course the $a^{\prime 3}$ result follows from multiplying the $\alpha$
integrand by $1/a\alpha$.

As might be expected, the reduction of $f_6(x)$ is the most complicated.
The factor of $x \ln^2(x)$ times the $a^{\prime 4}$ measure factor gives,
\begin{equation}
x \ln^2(x) \Longrightarrow \frac{16 \pi^2 i}{H^4} \int_{\frac1{a}}^1 
\frac{d\alpha}{\alpha^2} (1 \!-\! \alpha)^5 \Biggl\{\frac1{30} \ln(\alpha) 
\!-\! \frac1{15} \ln(1 \!-\! \alpha) \!+\! \frac{31}{450} \Biggr\} \; .
\end{equation}
The second factor produces,
\begin{eqnarray}
\lefteqn{-2 x \ln(1 \!-\! x) \ln(x) \Longrightarrow \frac{16 \pi^2 i}{H^4} 
\int_{\frac1{a}}^1 \frac{d\alpha}{\alpha^2} \Biggl\{\Bigl[-\frac1{30} \!+\!
\frac{\alpha}6 \!+\! \frac{2 \alpha^2}{3} \!+\! \frac{2 \alpha^3}{3} \!+\!
\frac{\alpha^4}6 \!-\! \frac{\alpha^5}{30}\Bigr] \ln(\alpha) } \nonumber \\
& & \hspace{5cm} - \frac{31}{450} \!+\! \frac{11\alpha}{18} \!+\! 
\frac{\alpha^2}{9} \!-\! \frac{\alpha^3}{9} \!-\! \frac{11\alpha^4}{18} 
\!+\! \frac{31 \alpha^5}{450} \Biggr\} . \qquad
\end{eqnarray}
And the final term contributes,
\begin{equation}
-2 \ln(x) \Longrightarrow \frac{16 \pi^2 i}{H^4} \int_{\frac1{a}}^1 
\frac{d\alpha}{\alpha^2} \times -\frac{\alpha}3 (1 \!-\! \alpha)^3 \; .
\end{equation}
The sum of all three terms is,
\begin{eqnarray}
\lefteqn{\frac{16 \pi^2 i}{H^4} \int_{\frac1{a}}^1 d\alpha \Biggl\{
\Bigl[1 \!+\! \frac{\alpha}3 \!+\! \frac{\alpha^2}{3} \!-\! \frac{\alpha^3}{15}
\Bigr] \ln(\alpha) } \nonumber \\
& & \hspace{3.5cm} -\frac1{15 \alpha^2} (1 \!-\! \alpha)^5 \ln(1\!-\! \alpha)
\!-\! \frac1{15 \alpha} \!+\! \frac95 \!-\! \frac{9 \alpha}5 \!+\! 
\frac{\alpha^2}{15} \Biggr\} \; , \qquad \\
& & = \frac{16 \pi^2 i}{H^4} \Biggl\{ \frac5{24} - \frac{\pi^2}{18} + 
O\Bigl(\frac{\ln(a)}{a}\Bigr) \Biggr\} \; . 
\end{eqnarray}
As always, one obtains the result for the $a^{\prime 3}$ measure factor from
multiplying the $\alpha$ integrand by $1/a\alpha$.

\centerline{\bf Acknowledgements}

This work was partially supported by the Institute for Theoretical Physics
of Utrecht University, by the European Social fund and National resources 
$\Upsilon\Pi{\rm E}\Pi\Theta$-PythagorasII-2103, by European Union grants 
MRTN-CT-2004-512194 and FP-6-12679, by NSF grant PHY-0244714, and by the 
Institute for Fundamental Theory at the University of Florida.


\begin{thebibliography}{99}

\bibitem{RPW1} R. P. Woodard, ``Quantum Effects during Inflation,'' in
{\it Norman 2003, Quantum field theory under the influence of external
conditions} (Rinton Press, Princeton, 2004) ed. K. A. Milton, pp. 325-330,
astro-ph/0310757.

\bibitem{VF} A. Vilenkin and L. H. Ford, Phys. Rev. {\bf D26} (1982) 1231.

\bibitem{L} A. D. Linde, Phys. Lett. {\bf B116} (1982) 335.

\bibitem{S} A. A. Starobinski\u{\i}, Phys. Lett. {\bf B117} (1982) 175.

\bibitem{OW1} V. K. Onemli and R. P. Woodard, Class. Quant. Grav. {\bf 19}
(2002) 4607, gr-qc/0204065.

\bibitem{OW2} V. K. Onemli and R. P. Woodard, Phys. Rev. {\bf D70} (2004) 
107301, gr-qc/0406098.

\bibitem{BOW} T. Brunier, V. K. Onemli and R. P. Woodard, Class. Quant. Grav. 
{\bf 22} (2005) 59, gr-qc/0408080.

\bibitem{PTW1} T. Prokopec, O. Tornkvist and R. P. Woodard, Phys. Rev. Lett.
{\bf 89} (2002) 101301, astro-ph/0205331.

\bibitem{PTW2} T. Prokopec, O. Tornkvist and R. P. Woodard, Ann. Phys.
{\bf 303} (2003) 251, gr-qc/0205130.

\bibitem{PW1} T. Prokopec and R. P. Woodard, Am. J. Phys. {\bf 72} (2004) 60, 
astro-ph/0303358.

\bibitem{PW2} T. Prokopec and R. P. Woodard, Ann. Phys. {\bf 312} (2004) 1, 
gr-qc/0310056.

\bibitem{PW3} T. Prokopec and R. P. Woodard, JHEP {\bf 0310} (2003) 059, 
astro-ph/0309593.

\bibitem{GP} B. Garbrecht and T. Prokopec, Phys. Rev. {\bf D73} (2006)
064036, gr-qc/0602011.

\bibitem{TW1} N. C. Tsamis and R. P. Woodard, Ann. Phys. {\bf 238} (1995) 1.

\bibitem{TW2} N. C. Tsamis and R. P. Woodard, Nucl. Phys. {\bf B474} (1996) 
235, hep-ph/9602315.
 
\bibitem{TW3} N. C. Tsamis and R. P. Woodard, Ann. Phys. {\bf 253} (1997) 1,
hep-ph/9602316.

\bibitem{MW1} S. P. Miao and R. P. Woodard, Class. Quant. Grav. {\bf 23}
(2006) 1721, gr-qc/0511140.

\bibitem{MW2} S. P. Miao and R. P. Woodard, Phys. Rev. {\bf D74} (2006)
024021, gr-qc/0603135.

\bibitem{SW1} S. Weinberg, Phys. Rev. {\bf D72} (2005) 043514, hep-th/0506236.

\bibitem{SW2} S. Weinberg, Phys. Rev. {\bf D74} (2006) 023508, hep-th/0605244.

\bibitem{AAS} A. A. Starobinski\u{\i}, ``Stochastic de Sitter (inflationary)
stage in the early universe,'' in {\it Field Theory, Quantum Gravity and
Strings}, ed. H. J. de Vega and N. Sanchez (Springer-Verlag, Berlin, 1986)
pp. 107-126.

\bibitem{AV} A. Vilenkin, Phys. Rev. {\bf D27} (1983) 2848.

\bibitem{NS} Y. Nambu and M. Sasaki, Phys. Lett. {\bf 219} (1989) 240.

\bibitem{GLM} A. S. Goncharov, A. D. Linde and V. F. Mukhanov, Int. J.
Mod. Phys. {\bf A2} (1987) 561.

\bibitem{LM} A. D. Linde and A. Mezhlumian, Phys. Lett. {\bf B307} (1993)
25, gr-qc/9304015.

\bibitem{SJR} S. J. Rey, Nucl. Phys. {\bf B284} (1987) 706.

\bibitem{SNN} M. Sasaki, Y. Nambu and K. I. Nakao, Nucl. Phys. {\bf B308}
(1988) 868.

\bibitem{WV} S. Winitzki and A. Vilenkin, Phys. Rev. {\bf D61} (2000)
084008, gr-qc/9911029.

\bibitem{RPW2} R. P. Woodard, Nucl. Phys. Proc. Suppl. {\bf 148} (2005), 108
astro-ph/0502556.

\bibitem{TW4} N. C. Tsamis and R. P. Woodard, Nucl. Phys. {\bf B724} (2005)
295, gr-qc/0505115.

\bibitem{SY} A. A. Starobinski\u{\i} and J. Yokoyama, Phys. Rev. {\bf D50}
(1994) 6357, astro-ph/9407016.

\bibitem{MW3} S. P. Miao and R. P. Woodard, Phys. Rev. {\bf D74} (2006)
044019, gr-qc/0602110.

\bibitem{PTsW1} T. Prokopec, N. C. Tsamis and R. P. Woodard, ``Stochastic
Inflationary Scalar Electrodynamics,'' arXiv:0707.0847.

\bibitem{AJ} B. Allen and T. Jacobson, Commun. Math. Phys. {\bf 103} (1987)
669.

\bibitem{KW} E. O. Kahya and R. P. Woodard, Phys. Rev. {\bf D72} (2205) 104001,
gr-qc/0508015.

\bibitem{AF} B. Allen and A. Folacci, Phys. Rev. {\bf D35} (1987) 3771.

\bibitem{BA} B. Allen, Phys. Rev. {\bf D32} (1985) 3136.

\bibitem{TW5} N. C. Tsamis and R. P. Woodard, J. Math. Phys. {\bf 48} (2007)
052306, gr-qc/0608069.

\bibitem{EW} E. Witten, ``Quantum gravity in de Sitter space,'' in
{\it New Fields and Strings in Subnuclear Physics: Proceedings of the
International School of Subnuclear Physics (Subnuclear Series, 39)},
ed. A Zichichi (World Scientific, Singapore, 2002), hep-th/0106109.

\bibitem{AS} A. Strominger, JHEP {\bf 0110} (2001) 034, hep-th/0106113.

\bibitem{TW6} N. C. Tsamis and R. P. Woodard, Class. Quant. Grav. {\bf 11}
(1994) 2969.

\bibitem{JS} J. Schwinger, J. Math. Phys. {\bf 2} (1961) 407.

\bibitem{M} K. T. Mahanthappa, Phys. Rev. {\bf 126} (1962) 329.

\bibitem{BM} P. M. Bakshi and K. T. Mahanthappa, J. Math. Phys. {\bf 4} (1963)
1; J. Math. Phys. {\bf 4} (1963) 12.

\bibitem{K} L. V. Keldysh, Sov. Phys. JETP {\bf 20} (1965) 1018.

\bibitem{CSHY} K. C. Chou, Z. B. Su, B. L. Hao and L. Yu, Phys. Rept. {\bf 118} 
(1985) 1.

\bibitem{J} R. D. Jordan, Phys. Rev. {\bf D33} (1986) 444.

\bibitem{CH} E. Calzetta and B. L. Hu, Phys. Rev. {\bf D35} (1987) 495.

\bibitem{FW} L. H. Ford and R. P. Woodard, Class. Quant. Grav. {\bf 22} (2005)
1637, gr-qc/0411003.

\bibitem{TW7} N. C. Tsamis and R. P. Woodard, Ann. Phys. {\bf 238} (1995) 1.

\bibitem{TW8} N. C. Tsamis and R. P. Woodard, Class. Quant. Grav. {\bf 2}
(1985) 841.

\bibitem{KW2} E. O. Kahya and R. P. Woodard, Phys. Rev. {\bf D74} (2006)
084012, gr-qc/0608049.

\bibitem{DW} L. D. Duffy and R. P. Woodard, Phys. Rev. {\bf D72} (2005) 
024023, hep-ph/0505156.

\end{thebibliography}
\end{document}